\documentclass[times,twocolumn]{aastex62}

\usepackage{cases}
\usepackage{graphicx}
\usepackage{subfigure}
\usepackage{natbib}
\usepackage{color}
\usepackage{tabularx}
\usepackage{bm}
\usepackage{threeparttable}
\newcommand{\thetae}{\theta_{\rm E}}

\newcommand{\pie}{{\pi}_{\rm E}}

\newcommand{\te}{t_{\rm E}}
\newcommand{\event}{OGLE-2017-BLG-0448}

\newcommand{\hjd}{{\rm HJD}^{\prime}}
\definecolor{tricolor}{rgb}{1,0,0}
\DeclareGraphicsExtensions{.pdf,.png,.jpg}

\shorttitle{}
\shortauthors{Zhai et al.}

\begin{document}
\title{{\large OGLE-2017-BLG-0448Lb: A Low Mass-Ratio Wide-Orbit Microlensing Planet?}}

%\correspondingauthor{Radoslaw Poleski, Weicheng Zang}
%\email{rpoleski@astrouw.edu.pl, 3130102785@zju.edu.cn}

\author[0009-0004-1650-3494]{Ruocheng Zhai}
\affiliation{Department of Astronomy, Tsinghua University, Beijing 100084, China}

\author[0000-0002-9245-6368]{Rados\l{}aw Poleski}
\affiliation{Astronomical Observatory, University of Warsaw, Al. Ujazdowskie 4, 00-478 Warszawa, Poland}
%rpoleski@astrouw.edu.pl

\author[0000-0001-6000-3463]{Weicheng Zang}
\affiliation{Center for Astrophysics $|$ Harvard \& Smithsonian, 60 Garden St.,Cambridge, MA 02138, USA}
\affiliation{Department of Astronomy, Tsinghua University, Beijing 100084, China}

\author{Youn Kil Jung}
\affiliation{Korea Astronomy and Space Science Institute, Daejon 34055, Republic of Korea}
\affiliation{University of Science and Technology, Korea, (UST), 217 Gajeong-ro Yuseong-gu, Daejeon 34113, Republic of Korea}
%ykjung21@kasi.re.kr

\author[0000-0001-5207-5619]{Andrzej Udalski}
\affiliation{Astronomical Observatory, University of Warsaw, Al. Ujazdowskie 4, 00-478 Warszawa, Poland}
%udalski@astrouw.edu.pl

\author[0000-0003-2337-0533]{Renkun Kuang}
\affiliation{Department of Astronomy, Tsinghua University, Beijing 100084, China}
\affiliation{Department of Engineering Physics, Tsinghua University, Beijing 100084, China}

\collaboration{(Leading Authors)}

\author{Michael D. Albrow}
\affiliation{University of Canterbury, Department of Physics and Astronomy, Private Bag 4800, Christchurch 8020, New Zealand}
%michael.albrow@canterbury.ac.nz

\author{Sun-Ju Chung}
\affiliation{Korea Astronomy and Space Science Institute, Daejon 34055, Republic of Korea}
%sjchung@kasi.re.kr

\author{Andrew Gould}
\affiliation{Max-Planck-Institute for Astronomy, K\"onigstuhl 17, 69117 Heidelberg, Germany}
\affiliation{Department of Astronomy, Ohio State University, 140 W. 18th Ave., Columbus, OH 43210, USA}
%andygould47@yahoo.com

\author{Cheongho Han}
\affiliation{Department of Physics, Chungbuk National University, Cheongju 28644, Republic of Korea}
%cheongho@astroph.chungbuk.ac.kr

\author{Kyu-Ha Hwang}
\affiliation{Korea Astronomy and Space Science Institute, Daejon 34055, Republic of Korea}
%kyuha@kasi.re.kr

\author{Yoon-Hyun Ryu}
\affiliation{Korea Astronomy and Space Science Institute, Daejon 34055, Republic of Korea}
%yoonhyunryu@gmail.com

\author{In-Gu Shin}
\affiliation{Center for Astrophysics $|$ Harvard \& Smithsonian, 60 Garden St.,Cambridge, MA 02138, USA}
%ingushin@gmail.com

\author{Yossi Shvartzvald}
\affiliation{Department of Particle Physics and Astrophysics, Weizmann Institute of Science, Rehovot 76100, Israel}
%yossishv@gmail.com

\author[0000-0003-0626-8465]{Hongjing Yang}
\affiliation{Department of Astronomy, Tsinghua University, Beijing 100084, China}
%Yang-hj19@mails.tsinghua.edu.cn

\author{Jennifer C. Yee}
\affiliation{Center for Astrophysics $|$ Harvard \& Smithsonian, 60 Garden St.,Cambridge, MA 02138, USA}
%jyee@cfa.harvard.edu

\author{Sang-Mok Cha}
\affiliation{Korea Astronomy and Space Science Institute, Daejon 34055, Republic of Korea}
\affiliation{School of Space Research, Kyung Hee University, Yongin, Kyeonggi 17104, Republic of Korea} 
%chasm@kasi.re.kr

\author{Dong-Jin Kim}
\affiliation{Korea Astronomy and Space Science Institute, Daejon 34055, Republic of Korea}
%keaton03@kasi.re.kr

\author{Hyoun-Woo Kim}
\affiliation{Korea Astronomy and Space Science Institute, Daejon 34055, Republic of Korea}
%hwkim@kasi.re.kr

\author{Seung-Lee Kim}
\affiliation{Korea Astronomy and Space Science Institute, Daejon 34055, Republic of Korea}
%slkim@kasi.re.kr

\author{Chung-Uk Lee}
\affiliation{Korea Astronomy and Space Science Institute, Daejon 34055, Republic of Korea}
%leecu@kasi.re.kr

\author{Dong-Joo Lee}
\affiliation{Korea Astronomy and Space Science Institute, Daejon 34055, Republic of Korea}
%marin678@kasi.re.kr

\author{Yongseok Lee}
\affiliation{Korea Astronomy and Space Science Institute, Daejon 34055, Republic of Korea}
\affiliation{School of Space Research, Kyung Hee University, Yongin, Kyeonggi 17104, Republic of Korea}
%yslee@kasi.re.kr

\author{Byeong-Gon Park}
\affiliation{Korea Astronomy and Space Science Institute, Daejon 34055, Republic of Korea}
\affiliation{University of Science and Technology, Korea, (UST), 217 Gajeong-ro Yuseong-gu, Daejeon 34113, Republic of Korea}
%bgpark@kasi.re.kr

\author{Richard W. Pogge}
\affiliation{Department of Astronomy, Ohio State University, 140 W. 18th Ave., Columbus, OH  43210, USA}
%pogge.1@osu.edu

\collaboration{(The KMTNet Collaboration)}

\author[0000-0002-2335-1730]{Jan Skowron}
\affiliation{Astronomical Observatory, University of Warsaw, Al. Ujazdowskie 4, 00-478 Warszawa, Poland}
%jskowron@astrouw.edu.pl

\author[0000-0002-0548-8995]{Micha{\l}~K. Szyma\'{n}ski}
\affiliation{Astronomical Observatory, University of Warsaw, Al. Ujazdowskie 4, 00-478 Warszawa, Poland}
%msz@astrouw.edu.pl

\author[0000-0002-7777-0842]{Igor Soszy\'{n}ski}
\affiliation{Astronomical Observatory, University of Warsaw, Al. Ujazdowskie 4, 00-478 Warszawa, Poland}
%soszynsk@astrouw.edu.pl

\author[0000-0001-6364-408X]{Krzysztof Ulaczyk}
\affiliation{Department of Physics, University of Warwick, Gibbet Hill Road, Coventry, CV4~7AL,~UK}
%kulaczyk@astrouw.edu.pl

\author[0000-0002-2339-5899]{Pawe{\l} Pietrukowicz}
\affiliation{Astronomical Observatory, University of Warsaw, Al. Ujazdowskie 4, 00-478 Warszawa, Poland}
%pietruk@astrouw.edu.pl

\author[0000-0003-4084-880X]{Szymon Koz{\l}owski}
\affiliation{Astronomical Observatory, University of Warsaw, Al. Ujazdowskie 4, 00-478 Warszawa, Poland}
%simkoz@astrouw.edu.pl

\author[0000-0001-7016-1692]{Przemek Mr\'{o}z}
\affiliation{Astronomical Observatory, University of Warsaw, Al. Ujazdowskie 4, 00-478 Warszawa, Poland}
%pmroz@astrouw.edu.pl

\author[0000-0002-9326-9329]{Krzysztof A. Rybicki}
\affiliation{Department of Particle Physics and Astrophysics, Weizmann Institute of Science, Rehovot 76100, Israel}
\affiliation{Astronomical Observatory, University of Warsaw, Al. Ujazdowskie 4, 00-478 Warszawa, Poland}
%krybicki@astrouw.edu.pl

\author[0000-0002-6212-7221]{Patryk Iwanek}
\affiliation{Astronomical Observatory, University of Warsaw, Al. Ujazdowskie 4, 00-478 Warszawa, Poland}
%piwanek@astrouw.edu.pl

\author[0000-0002-3051-274X]{Marcin Wrona}
\affiliation{Astronomical Observatory, University of Warsaw, Al. Ujazdowskie 4, 00-478 Warszawa, Poland}
%mwrona@astrouw.edu.pl

\author[0000-0002-1650-1518]{Mariusz Gromadzki}
\affiliation{Astronomical Observatory, University of Warsaw, Al. Ujazdowskie 4, 00-478 Warszawa, Poland}
%Mazki@astrouw.edu.pl

\collaboration{(The OGLE Collaboration)}

%shude.mao@gmail.com

\author{Hanyue Wang}
\affiliation{Center for Astrophysics $|$ Harvard \& Smithsonian, 60 Garden St.,Cambridge, MA 02138, USA}

\author{Shude Mao}
\affiliation{Department of Astronomy, Tsinghua University, Beijing 100084, China}

\author{Jiyuan Zhang}
\affiliation{Department of Astronomy, Tsinghua University, Beijing 100084, China}

\author{Qiyue Qian}
\affiliation{Department of Astronomy, Tsinghua University, Beijing 100084, China}

\author[0000-0003-4027-4711]{Wei Zhu}
\affiliation{Department of Astronomy, Tsinghua University, Beijing 100084, China}
%weizhu@mail.tsinghua.edu.cn

\collaboration{(The MAP Collaboration)}

\begin{abstract}

The gravitational microlensing technique is most sensitive to planets in a Jupiter-like orbit and has detected more than 200 planets. However, only a few wide-orbit ($s > 2$) microlensing planets have been discovered, where $s$ is the planet-to-host separation normalized to the angular Einstein ring radius, $\thetae$. Here we present the discovery and analysis of a strong candidate wide-orbit microlensing planet in the event, OGLE-2017-BLG-0448. The whole light curve exhibits long-term residuals to the static binary-lens single-source model, so we investigate the residuals by adding the microlensing parallax, microlensing xallarap, an additional lens, or an additional source. For the first time, we observe a complex degeneracy between all four effects. The wide-orbit models with $s \sim 2.5$ and a planet-to-host mass-ratio of $q \sim 10^{-4}$ are significantly preferred, but we cannot rule out the close models with $s \sim 0.35$ and $q \sim 10^{-3}$. A Bayesian analysis based on a Galactic model indicates that, despite the complicated degeneracy, the surviving wide-orbit models all contain a super-Earth-mass to Neptune-mass planet at a projected planet-host separation of $\sim 6$ au and the surviving close-orbit models all consist of a Jovian-mass planet at $\sim 1$ au. The host star is probably an M or K dwarf. 
We discuss the implications of this dimension-degeneracy disaster on microlensing light-curve analysis and its potential impact on statistical studies.
%\textbf{Our investigations on the additional four effects lead to several concerns for microlensing studies, including measurements of the annual microlensing parallax, statistical studies of stellar binaries, and dimension-degeneracy disasters in microlensing light-curve analysis.}

\end{abstract}

\section{Introduction}\label{intro}
The Solar System planets are typically divided into three groups: rocky planets, gas giants, and ice giants. The two groups of giant planets are more important than the rocky planets from the perspective of planetary system formation and evolution: the amount of water on Earth is influenced by the time when Jupiter's core formed \citep{2016Icar..267..368M}, the changes of orbits of most massive planets significantly changed orbits of other planets and dwarf planets \citep{1999Natur.402..635T,2005Natur.435..459T, 2016AJ....151...22B} even leading to an ejection of a planet \citep{2012ApJ...744L...3B,2012AJ....144..117N}, to name just a few aspects.

In the Solar System all giant planets have orbits wider than the ice-line \citep[$2.7~\mathrm{AU}$;][]{snowline}. Hence, in order to understand the Solar System formation in a broader context, we should be interested in searching for exoplanets orbiting other stars on similarly wide orbits. Currently, there are only two exoplanet detection techniques that efficiently find planets on wide orbits: direct imaging and gravitational microlensing \citep{Shude1991,Andy1992}. Microlensing has a unique capability to find wide-orbit exoplanets with mass ratios of Jupiter to the Sun ($10^{-3}$) or lower \citep{Gaudi2012}. The two planet parameters that are routinely measured are the mass ratio ($q$) and projected separation ($s$), which is measured relative to the angular Einstein ring radius ($\thetae$). In a typical case, the Einstein ring radius corresponds to the projected planet-star separations on the order of $2.5~\mathrm{AU}$. Hence, to study exoplanets on orbits similar to the Solar System giant planets, we should focus on microlensing exoplanets that have $s \gg 1$.

Here we present a detailed analysis of \event Lb, which is a strong candidate for a wide-orbit planet with a low mass ratio. This wide orbit solution can be compared to the widest-orbit microlensing planet: OGLE-2008-BLG-092LAb with $s=5.26\pm0.11$ and $q=(2.41\pm0.45)\times10^{-4}$ \citep{OB08092}. Recent microlensing studies have focused on planets with mass ratios of $10^{-4}$ and smaller because of possible break of the mass-ratio distribution function \citep{Suzuki2016, OB171434, KB170165}. Among the $q<10^{-4}$ planets, the widest secure separation is $s=1.610\pm0.008$ for OGLE-2005-BLG-390Lb \citep[$q=(0.76\pm0.07)\times10^{-4}$;][]{OB05390}. A larger separation of $s=1.773\pm0.006$ (and $q=(0.187\pm0.015)\times 10^{-4}$) is possible for OGLE-2018-BLG-0596Lb but the light curve of this planet favors the close solution ($s=0.564\pm0.005$ and $q=(1.33\pm0.11)\times 10^{-4}$) by $\Delta\chi^2 = 17$ \citep{OB180596}. The wide solution for \event Lb has $q$ smaller by a factor of 
$6.7$ than OGLE-2008-BLG-092LAb and separation wider by a factor of $\gtrsim 1.5$ % XXX - CHECK THESE NUMBERS AT THE END.
than OGLE-2005-BLG-390Lb or OGLE-2018-BLG-0596Lb. Hence, \event Lb is unique in probing the mass-ratio distribution at the wide separations.

The detection of the planetary anomaly in this event was first mentioned by \citet{-4planet} who presented a systematic search \citep{OB191053,2019_prime} for planets in the Korean Microlensing Telescope Network \citep[KMTNet;][]{KMT2016} photometric database from 2016--2019. However, the complexity of the event analysis required a detailed investigation presented here.

\section{Observations}\label{obser}

The source of the microlensing event \event lies toward the Galactic bulge at the equatorial coordinates $(\alpha, \delta)_{\rm J2000}$ = (17:54:40.47, $-31$:01:54.9), corresponding to Galactic coordinates $(\ell,b)=(-0.7948, -2.7576)$. The event was found first by the Optical Gravitational Lensing Experiment (OGLE) and announced on 31 March 2017 by the OGLE Early Warning System \citep{Udalski2003,OGLEIV}. The event was then independently discovered by the KMTNet post-season EventFinder system \citep{KMTeventfinder} based on all the data collected during the 2017 season. 

The OGLE survey obtains photometry using a 1.3m telescope with 1.4 ${\rm deg}^2$ camera at the Las Campanas Observatory \citep[Chile;][]{OGLEIV}. The event was located in the OGLE field BLG534, which is observed with a cadence of $1~{\rm hr}^{-1}$. The KMTNet survey conducted observations from three identical 1.6 m telescopes equipped with $4~{\rm deg}^2$ cameras in Chile (KMTC), South Africa (KMTS), and Australia (KMTA). The event lies in two slightly offset KMT fields, BLG01 and BLG41, with a combined cadence of $4~{\rm hr}^{-1}$ for KMTC and $3~{\rm hr}^{-1}$ for KMTA and KMTS. For both surveys, most of the images were taken in the $I$ band, and a small fraction of $V$-band images were acquired for source color measurements. For this event, the $V$-band data of KMTC41 and KMTS01 cover the planetary signal, so we include them in the light-curve analysis. The $V$-band photometry helps exclude the single-lens binary-source (1L2S) models.

The OGLE and KMTNet data used in the light-curve analysis were reduced using the custom photometry pipelines based on the difference imaging technique \citep{Tomaney1996,Alard1998}: pySIS \citep{pysis,Yang_TLC} for the KMTNet data, and \cite{Wozniak2000} for the OGLE data. For the KMTC01 data, we additionally conducted the pyDIA photometry \citep{pyDIA} to measure the source color and construct the color-magnitude diagram (CMD). The $I$-band magnitude of the light curve has been calibrated to the OGLE-III $I$-band magnitude \citep{OGLEIII}. The error bars for the OGLE and KMTNet data from the individual photometry pipelines were readjusted following the processes of \cite{OGLEerrorbar} and \cite{MB11293}, respectively. We note that using the \cite{OGLEerrorbar} method for the OGLE data leads to a 6\% difference between the total $\chi^2$ and the degree of freedom for the best-fit model, which does not influence the conclusion of our paper.

\begin{figure}[htb] 
    \centering
    \includegraphics[width=\columnwidth]{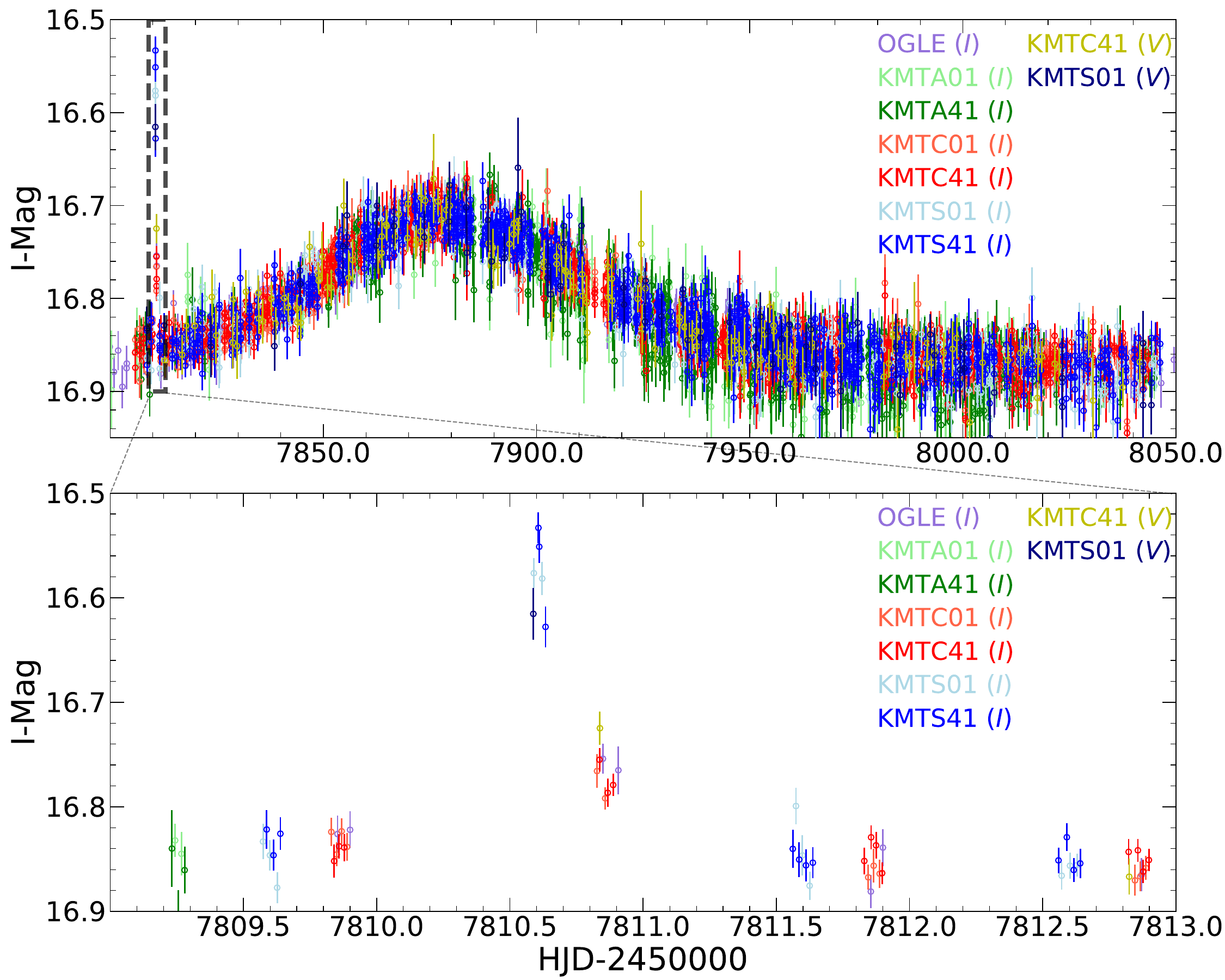}
    \caption{Observed light curve of the microlensing event, \event. Different colors represent the observed data from different data sets. The upper panel shows all of the data taken in 2017, and the low panel displays a close-up of the planetary signal.}
    \label{fig:lc1}
\end{figure}

\section{Binary-lens Single-source Model}\label{sec:2L1S}

\begin{table*}[htb]
\renewcommand\arraystretch{1.25}
\centering
\caption{Lensing Parameters for 2L1S Static Models}
\begin{tabular}{c|c c c c}
\hline
\hline
Parameters & Close Inner & Close Outer & Wide Inner & Wide Outer \\
\hline
$\chi^2$/dof & $\mathbf{10237.7/10687}$ & $10331.2/10687$ & $10281.1/10687$ & $10277.8/10687$ \\
\hline
$t_0$ & $\mathbf{7879.92_{-0.08}^{+0.08}}$ & $7879.99_{-0.09}^{+0.09}$ & $7879.95_{-0.09}^{+0.09}$ & $7879.94_{-0.09}^{+0.09}$ \\
$u_0$ & $\mathbf{1.225_{-0.031}^{+0.017}}$ & $1.225_{-0.031}^{+0.014}$ & $1.221_{-0.034}^{+0.017}$ & $1.224_{-0.030}^{+0.016}$ \\
$\te$ (days) & $\mathbf{32.51_{-0.39}^{+0.59}}$ & $32.39_{-0.32}^{+0.59}$ & $32.49_{-0.36}^{+0.65}$ & $32.45_{-0.34}^{+0.56}$ \\
$\rho$ ($10^{-2}$) & $\mathbf{0.078_{-0.035}^{+0.041}}$ & $0.294_{-0.051}^{+0.028}$ & $1.041_{-0.168}^{+0.074}$ & $0.665_{-0.378}^{+0.244}$ \\
$t_{\rm 0,pl}$ & $\mathbf{7813.08_{-0.13}^{+0.11}}$ & $7808.43_{-0.13}^{+0.12}$ & $7810.45_{-0.03}^{+0.04}$ & $7810.63_{-0.02}^{+0.02}$ \\
$u_{\rm 0,pl}$ & $\mathbf{0.123_{-0.006}^{+0.006}}$ & $-0.128_{-0.008}^{+0.008}$ & $0.005_{-0.003}^{+0.002}$ & $-0.007_{-0.001}^{+0.002}$ \\
$t_\mathrm{E, pl}$ (days) & $\mathbf{0.88_{-0.04}^{+0.04}}$ & $0.90_{-0.05}^{+0.06}$ & $0.18_{-0.02}^{+0.02}$ & $0.20_{-0.02}^{+0.01}$ \\
$I_{\rm S}$ & $\mathbf{17.30_{-0.04}^{+0.06}}$ & $17.30_{-0.03}^{+0.06}$ & $17.31_{-0.04}^{+0.07}$ & $17.30_{-0.03}^{+0.06}$ \\
\hline
$s$ & $\mathbf{0.3554_{-0.0033}^{+0.0054}}$ & $0.3545_{-0.0028}^{+0.0055}$ & $2.8198_{-0.0474}^{+0.0256}$ & $2.8141_{-0.0407}^{+0.0238}$ \\
$q$ ($10^{-4}$) & $\mathbf{7.384_{-0.672}^{+0.685}}$ & $7.623_{-0.881}^{+0.980}$ & $0.309_{-0.064}^{+0.088}$ & $0.362_{-0.067}^{+0.055}$ \\
$\alpha$ (deg) & $\mathbf{33.2_{-0.3}^{+0.2}}$ & $26.3_{-0.3}^{+0.3}$ & $209.8_{-0.2}^{+0.2}$ & $209.6_{-0.2}^{+0.2}$ \\
\hline
\hline
\end{tabular}
\label{tab:2L1S}
\tablecomments{We present the fitted and derived parameters above and below horizontal line, respectively. The best-fit 2L1S static model is boldfaced.}
\end{table*}

We display the light curve of the microlensing event, \event, in Figure \ref{fig:lc1}. The event started rising in early 2017. The first maximum of the brightness was observed at $\hjd = 7810$ (${\rm HJD}^{\prime} = {\rm HJD} - 2450000$). This maximum is mostly covered by the KMTS data. Its full amplitude is not precisely measured but it must be at least $0.25~\mathrm{mag}$. This maximum lasted one day or less, hence, was relatively short and we call it an anomaly henceforth. The exact shape of the anomaly is not well constrained because of a lack of data between $\mathrm{HJD'}=7809.9$ and $7810.6$. Following the anomaly, the event shows a long bell-shaped curve with an amplitude of $0.15~\mathrm{mag}$ and a peak at $\mathrm{HJD'}\approx7882$. A microlensing light curve with two maxima of similar shape can be interpreted as either a 1L2S event or a binary-lens single-source (2L1S) event \citep{Gaudi1998}. In the latter case, a large difference in the duration of the two maxima points to a low mass ratio, i.e., in the planetary regime \citep{GG1997}. Additionally, the two peaks have a relatively long time separation which points to a very wide ($s \gg 1$) or a very close ($s \ll 1$) lens topology \citep{Han2006}. We present the 2L1S analysis below and show the 1L2S analysis in Section \ref{sec:1L2S}.

\subsection{Static Binary-Lens Model}\label{sec:2L1S_static}

\begin{figure*}[ht] 
    \centering
    \includegraphics[width=.95\columnwidth]{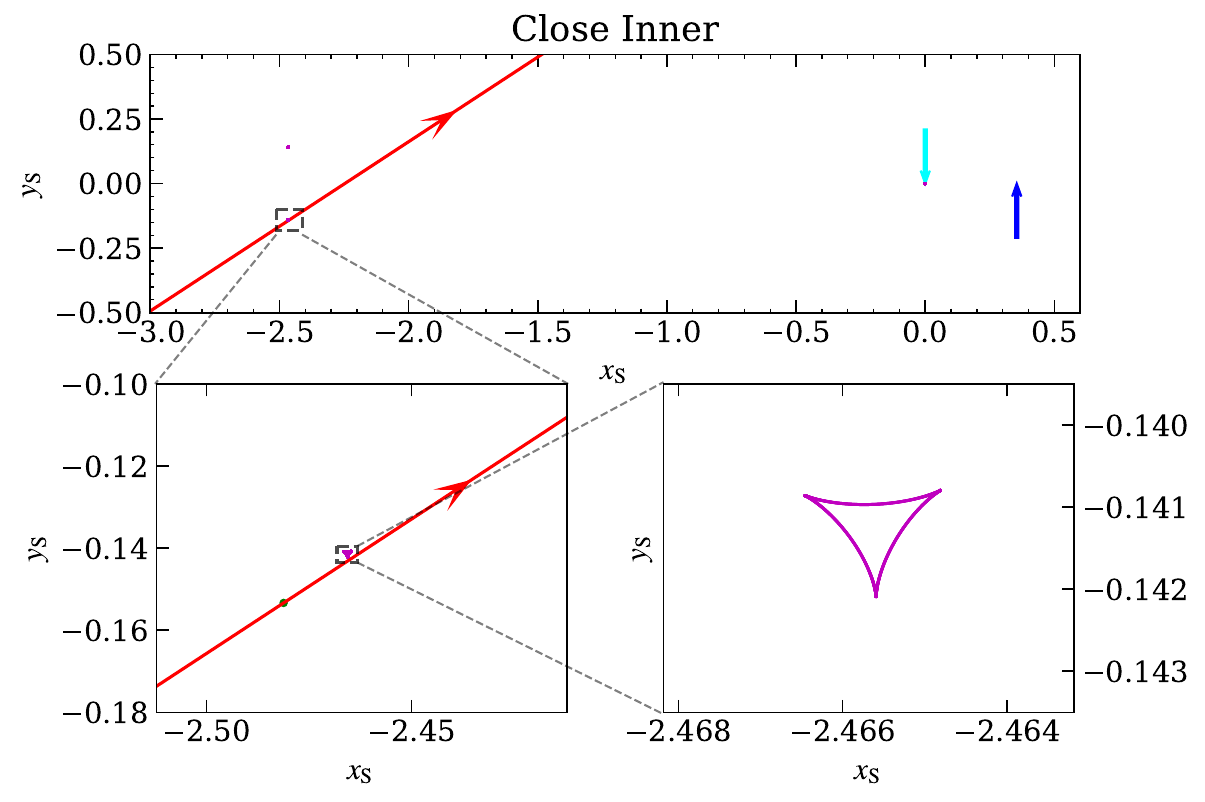}
    \includegraphics[width=.95\columnwidth]{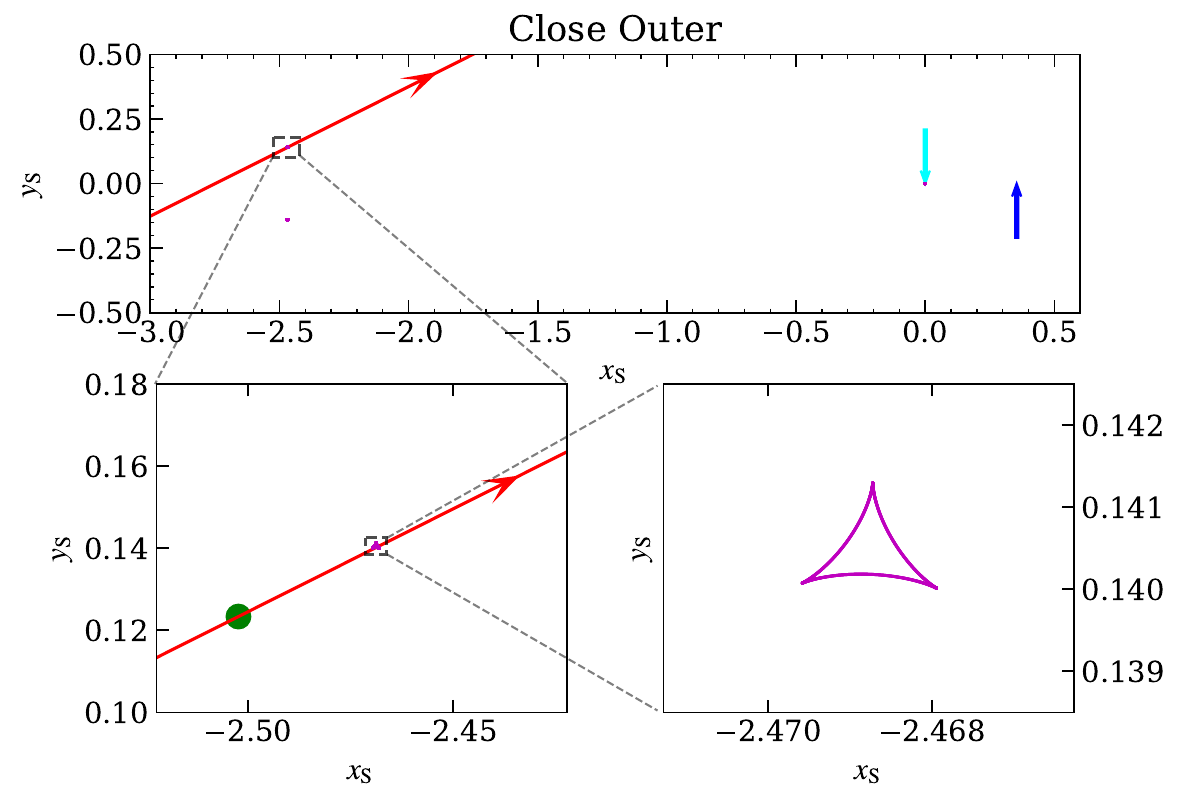}
    \includegraphics[width=.95\columnwidth]{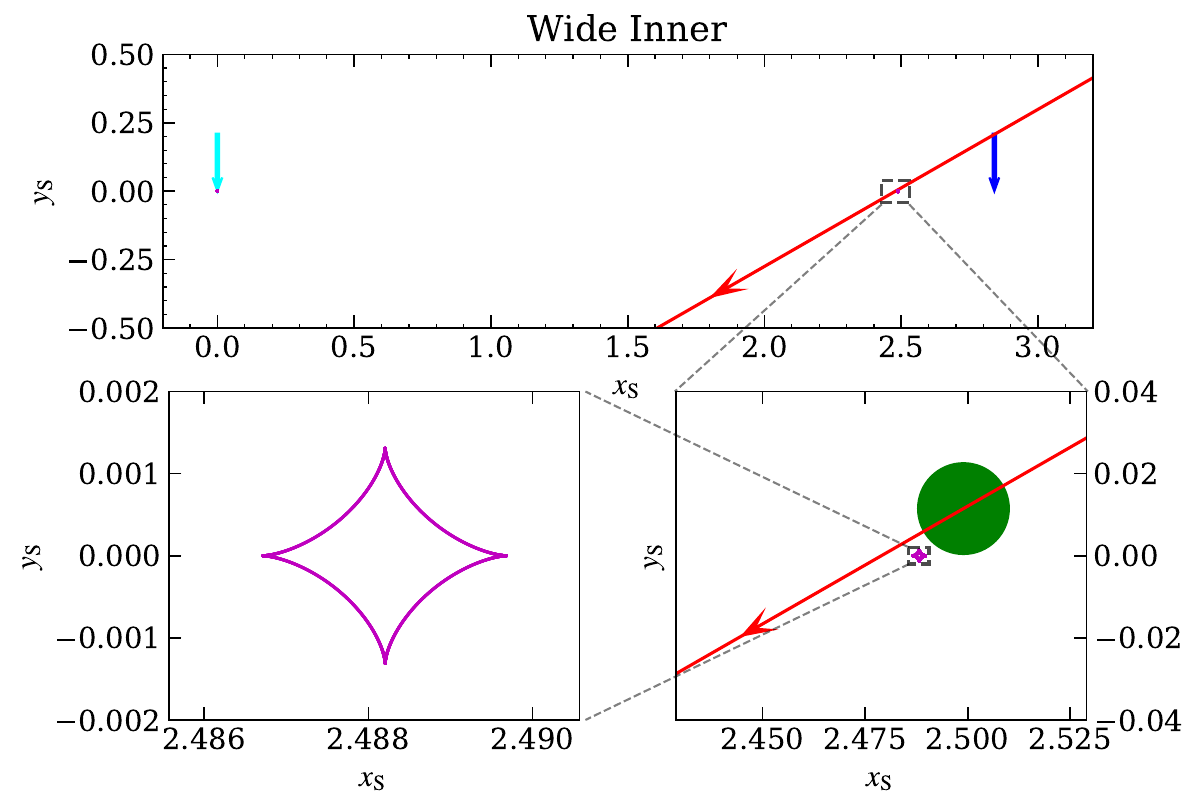}
    \includegraphics[width=.95\columnwidth]{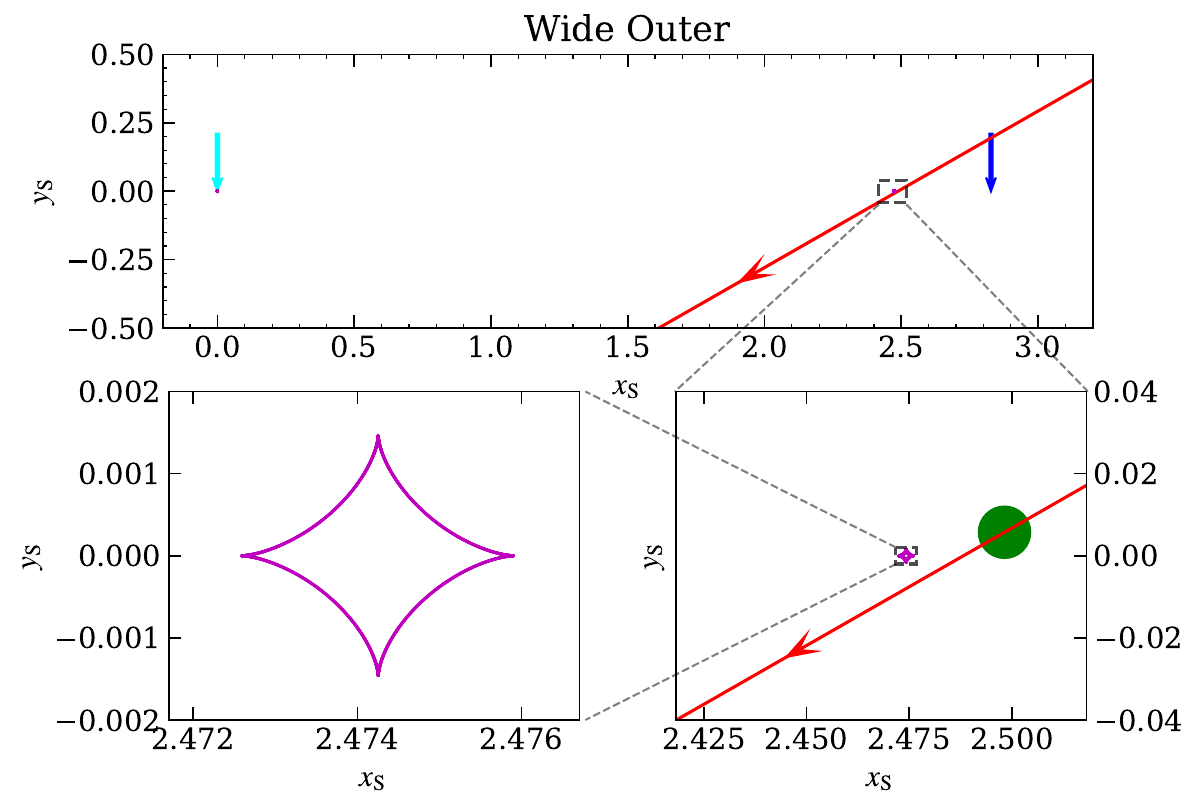}
    \caption{Caustic geometries of the four static 2L1S solutions. The locations of the host star and planet are indicated by cyan and blue arrows, respectively. The magenta lines show the caustic structures, the red lines with an arrow indicate the source trajectory and the direction of the source motion. The radii of the green dots represent the best-fit normalized source radius, $\rho$, of each solution.}
    \label{fig:cau1}
\end{figure*}

\begin{figure}[ht] 
    \centering
    \includegraphics[width=\columnwidth]{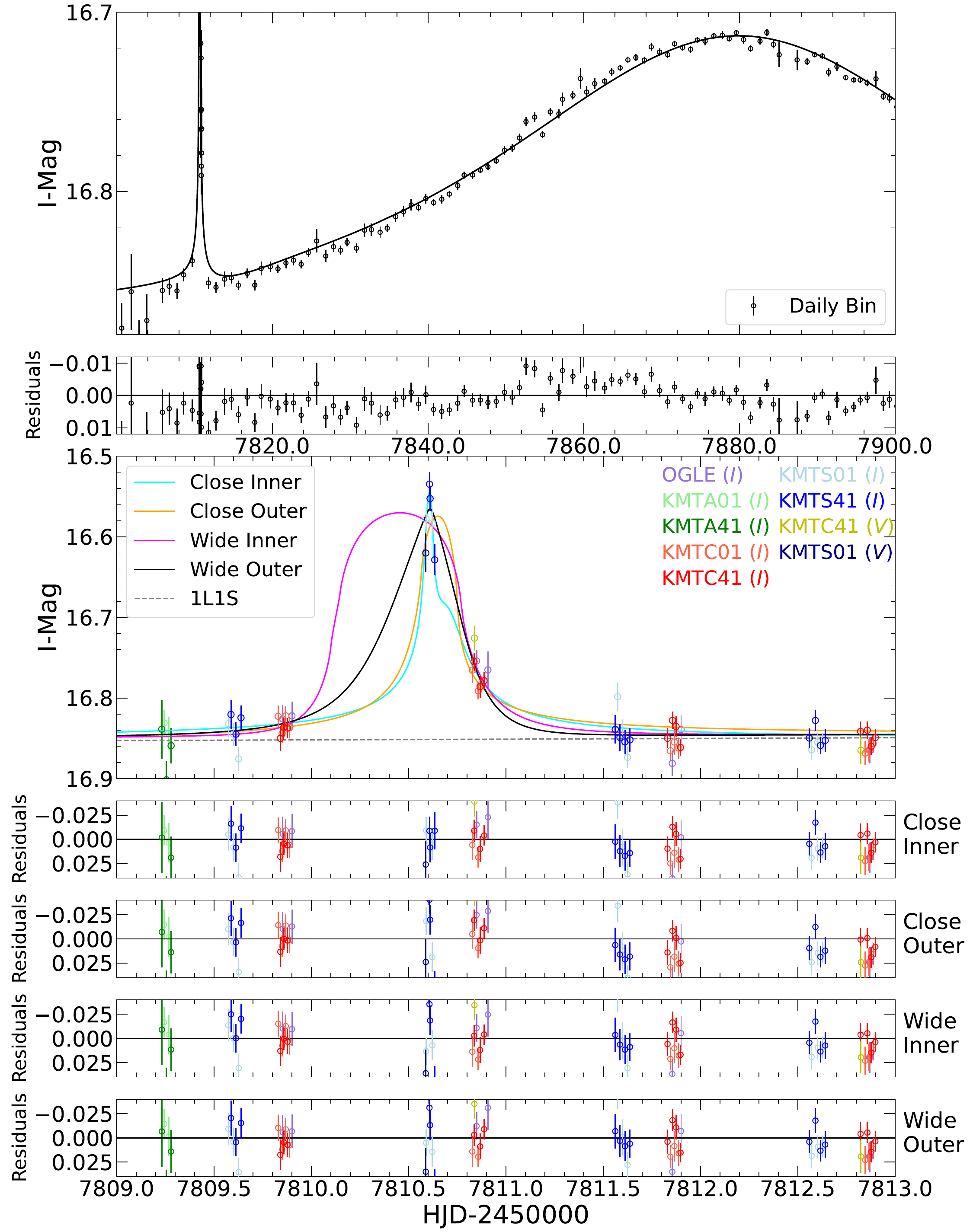}
    \caption{Observed data together with the four static 2L1S solutions. Data shown in the top two panels are daily binned. All four solutions can fit the anomaly well (the lower five panels), but there are long-term residuals before $\hjd = 7900$, so we include high-order effects.}
    \label{fig:lc2}
\end{figure}

The 2L1S model with a finite source is parametrized by seven variables. The first four are the same as for the single-lens single-source (1L1S, \citealt{Paczynski1986}) model:
$t_0$ -- the epoch of minimum lens--source separation,
$u_0$ -- the source--lens impact parameter relative to $\thetae$, 
$\te$ -- the Einstein ring crossing-time, and $\rho$ -- the ratio of the angular source size to $\thetae$. 
The other three parameters are typically: $s$, $q$, and $\alpha$ -- the angle between lens--source trajectory and the axis of the binary lens. In the case of \event, the epoch of the anomaly is well constrained by the data, and its duration is constrained to be short, but these two properties do not easily map on the $(s, q, \alpha)$ parameters. Hence, we decided to re-parametrize the model in order to improve convergence and acceptance ratios of the Markov chain Monte Carlo (MCMC) chains\footnote{We apply the \texttt{emcee} ensemble sampler \citep{emcee} for the MCMC $\chi^2$ minimization.}. Instead of $(s, q, \alpha)$ we use: 
$t_{0,\mathrm{pl}}$ -- the epoch of approach to the planetary caustic, 
$u_{0,\mathrm{pl}}$ -- the source--planetary caustic impact parameter relative to $\thetae$, and 
$t_\mathrm{E, pl}$ -- the planetary Einstein ring crossing time. We derive the equations to transform between the two sets of parameters based on a simple geometric consideration and the distance between planetary and central caustic \citep[$s' = \left|s-1/s\right|$;][]{Han2006}. For the wide topology, these equations are:
\begin{equation}
q = \left(\frac{t_{\mathrm{E, pl}}}{t_\mathrm{E}}\right)^2,
\end{equation}
\begin{equation}
\tau_{\rm pl} = \frac{t_0-t_{0,\mathrm{pl}}}{t_\mathrm{E}},
\end{equation}
\begin{equation}
s'=\sqrt{\left(u_0+u_{0,\mathrm{pl}}\right)^2+\tau_{\rm pl}^2},
\end{equation}
\begin{equation}
s = \frac{\sqrt{s'^2+4}+s'}{2},
\end{equation}
\begin{equation}
\alpha = \arcsin{\frac{u_0+u_{0,\mathrm{pl}}}{s'}}.
\end{equation}
For the close topology, the last two of these equations are modified:
\begin{equation}
s = \frac{\sqrt{s'^2+4}-s'}{2},
\end{equation}
\begin{equation}
\alpha = \arcsin{\frac{u_0+u_{0,\mathrm{pl}}}{s'}} + 180^\circ.
\end{equation}
%$$q = \left(\frac{t_{\mathrm{E, pl}}}{t_\mathrm{E}}\right)^2$$
%$$\tau = \frac{t_0-t_{0,\mathrm{pl}}}{t_\mathrm{E}}$$
%$$\eta = \tau\sin{\alpha} - u_0\cos{\alpha}$$
%$$s' = \frac{u_0 + \eta\cos{\alpha}}{\sin{\alpha}}$$
%$$s = \frac{\sqrt{s'^2+4}-s'}{2}$$
The above equations are also used for the binary lens models with additional higher-order effects: parallax and xallarap. We note that the physical interpretation of $t_{0,\mathrm{pl}}$, $u_{0,\mathrm{pl}}$, and $t_\mathrm{E, pl}$ provided above is only approximate for models with these higher-order effects.

We employ the advanced contour integration code \citep{Bozza2010, Bozza2018} to calculate the 2L1S magnification. In addition, we introduce two linear parameters ($f_{{\rm S},i}$, $f_{{\rm B},i}$) for each data set $i$ to represent the source flux and any blended flux. Both OGLE and KMTNet detected the event from the change in flux of a catalog star for which the OGLE-III catalog \citep{OGLEIII} gives the brightness of $I = 16.978 \pm 0.017$. Both surveys also reported a $>200~\mathrm{mas}$ offset between the magnified source and the catalog star. We thus check the $i'$-band baseline images taken by the 3.6m Canada-France-Hawaii Telescope (CFHT), with seeing FWHM of $0.''45$--$0.''50$. We calibrate the CFHT $i'$-band magnitude to the OGLE-III $I$-band magnitude using the field stars within $2'$ around the event. The CFHT images resolve two stars, with a brightness of $I = 17.29 \pm 0.02$ and $I = 18.11 \pm 0.07$, respectively, and the microlensing event occurred on the $I = 17.29 \pm 0.02$ star. Thus, we add the following prior to the blend flux $f_{\rm B}$: 
\begin{equation}
\mathcal{L}_{\rm prior} = \left\{\begin{array}{ll}
1 & {\rm if}\ f_{\rm B} \geq 0,\\
\exp(-\frac{f_{\rm B}^2}{2\sigma^2}) & {\rm if}\ f_{\rm B} < 0,
\end{array}\right.
\label{equ:blend}
\end{equation}
where $\sigma$ is the flux uncertainty of the $I = 17.29 \pm 0.02$ star. The posterior results are presented in Table \ref{tab:2L1S} with the MCMC fitting parameters presented first and additionally the $(s, q, \alpha)$ distributions provided at the end. We find four solutions, including two with minor-image (triangular, $s < 1$) planetary caustics and two with major-image (quadrilateral, $s > 1$) planetary caustics, as shown in Figure \ref{fig:cau1}. For every pair of solutions, the intersection between the source trajectory and the binary axis is either inside or outside the planetary caustics relative to the central caustic. Thus, we label the two $s < 1$ solutions as ``Close Inner'' and ``Close Outer'' and the two $s > 1$ solutions as ``Wide Inner'' and ``Wide Outer''. Figure \ref{fig:lc2} displays the light curves of the four solutions, and all of the four solutions can reasonably fit the data around the anomaly (i.e., the first maximum). However, as shown in the top two panels of Figure \ref{fig:lc2}, the static 2L1S model leaves long-term residuals before $\hjd = 7900$. Therefore, we further include high-order effects.

\begin{table*}[htb]
    \renewcommand\arraystretch{1.25}
    \centering
    \caption{Lensing Parameters for 2L1S Parallax Models}
    \begin{tabular}{c|c c|c c|c c}
    \hline
    \hline
    Parameters & \multicolumn{2}{c|}{Close Inner} & \multicolumn{2}{c|}{Close Outer} & \multicolumn{2}{c}{Wide} \\
     & $u_{0} > 0$ & $u_{0} < 0$ & $u_{0} > 0$ & $u_{0} < 0$ & $u_{0} > 0$ & $u_{0} < 0$ \\
    \hline
    $\chi^2$/dof & $10101.8/10683$ & $10116.7/10683$ & $10107.8/10683$ & $10119.7/10683$ & $\mathbf{10086.0/10683}$ & $10102.4/10683$ \\
    \hline
    $t_0$ & $7879.03_{-0.16}^{+0.13}$ & $7879.28_{-0.46}^{+0.13}$ & $7879.01_{-0.14}^{+0.11}$ & $7878.93_{-0.15}^{+0.15}$ & $\mathbf{7878.93_{-0.12}^{+0.12}}$ & $7879.15_{-0.17}^{+0.14}$ \\
    $u_0$ & $1.22_{-0.03}^{+0.02}$ & $-1.21_{-0.03}^{+0.05}$ & $1.17_{-0.12}^{+0.07}$ & $-1.23_{-0.01}^{+0.01}$ & $\mathbf{1.14_{-0.08}^{+0.07}}$ & $-1.22_{-0.02}^{+0.06}$ \\
    $\te$ (days) & $31.4_{-1.3}^{+0.9}$ & $27.4_{-0.9}^{+3.4}$ & $29.7_{-1.2}^{+2.0}$ & $32.5_{-0.8}^{+0.7}$ & $\mathbf{30.5_{-1.2}^{+1.6}}$ & $28.2_{-0.8}^{+1.6}$ \\
    $\rho$ ($10^{-2}$) & $0.13_{-0.05}^{+0.36}$ & $0.80_{-0.29}^{+0.11}$ & $0.67_{-0.06}^{+0.08}$ & $0.09_{-0.05}^{+0.20}$ & $\mathbf{1.81_{-0.33}^{+0.22}}$ & $1.37_{-0.69}^{+0.60}$ \\
    $\pi_{\rm E, N}$ & $0.36_{-0.23}^{+0.40}$ & $-2.19_{-0.24}^{+1.54}$ & $1.38_{-0.17}^{+0.14}$ & $-0.11_{-0.19}^{+0.15}$ & $\mathbf{1.19_{-0.14}^{+0.14}}$ & $-1.71_{-0.23}^{+0.42}$ \\
    $\pi_{\rm E, E}$ & $0.42_{-0.08}^{+0.16}$ & $0.36_{-0.05}^{+0.07}$ & $0.86_{-0.06}^{+0.07}$ & $0.34_{-0.03}^{+0.04}$ & $\mathbf{0.74_{-0.06}^{+0.05}}$ & $0.34_{-0.05}^{+0.04}$ \\
    $t_{\rm 0,pl}$ & $7810.6_{-2.7}^{+1.6}$ & $7814.0_{-4.9}^{+4.7}$ & $7807.4_{-4.0}^{+5.1}$ & $7813.8_{-1.8}^{+2.0}$ & $\mathbf{7799.5_{-4.7}^{+4.7}}$ & $7803.8_{-2.4}^{+2.7}$ \\
    $u_{\rm 0,pl}$ & $-0.09_{-0.19}^{+0.11}$ & $1.51_{-0.85}^{+0.29}$ & $-1.16_{-0.22}^{+0.27}$ & $0.00_{-0.06}^{+0.07}$ & $\mathbf{-0.78_{-0.14}^{+0.14}}$ & $1.04_{-0.23}^{+0.18}$ \\
    $t_\mathrm{E, pl}$ (days) & $1.36_{-0.14}^{+0.39}$ & $3.00_{-1.26}^{+0.43}$ & $3.02_{-0.42}^{+0.45}$ & $1.25_{-0.07}^{+0.11}$ & $\mathbf{0.37_{-0.04}^{+0.04}}$ & $0.36_{-0.05}^{+0.03}$ \\
    $ds/dt$ ($\rm yr^{-1}$) & $0.12_{-0.03}^{+0.06}$ & $0.38_{-0.16}^{+0.17}$ & $0.31_{-0.12}^{+0.17}$ & $0.21_{-0.05}^{+0.05}$ & $\mathbf{-0.32_{-1.12}^{+1.06}}$ & $-0.27_{-0.62}^{+0.67}$ \\
    $d\alpha/dt$ ($\rm yr^{-1}$) & $-0.4_{-5.9}^{+0.7}$ & $-44.5_{-64.9}^{+45.2}$ & $50.5_{-44.0}^{+55.6}$ & $0.2_{-0.5}^{+0.3}$ & $\mathbf{-0.1_{-21.6}^{+19.0}}$ & $-0.4_{-18.8}^{+15.6}$ \\
    $I_{\rm S}$ & $17.32_{-0.05}^{+0.06}$ & $17.34_{-0.06}^{+0.11}$ & $17.45_{-0.14}^{+0.26}$ & $17.29_{-0.02}^{+0.03}$ & $\mathbf{17.51_{-0.16}^{+0.18}}$ & $17.32_{-0.05}^{+0.12}$ \\
    \hline
    $\pi_{\rm E}$ & $0.55_{-0.18}^{+0.40}$ & $2.22_{-1.43}^{+0.23}$ & $1.63_{-0.18}^{+0.13}$ & $0.38_{-0.05}^{+0.09}$ & $\mathbf{1.40_{-0.14}^{+0.14}}$ & $1.74_{-0.39}^{+0.23}$ \\
    $s$ & $0.356_{-0.010}^{+0.007}$ & $0.365_{-0.020}^{+0.018}$ & $0.363_{-0.021}^{+0.022}$ & $0.368_{-0.008}^{+0.009}$ & $\mathbf{2.967_{-0.187}^{+0.168}}$ & $3.002_{-0.132}^{+0.120}$ \\
    $q$ ($10^{-4}$) & $18.64_{-4.16}^{+14.52}$ & $121.47_{-92.09}^{+36.51}$ & $104.17_{-26.53}^{+28.15}$ & $14.71_{-1.81}^{+3.24}$ & $\mathbf{1.46_{-0.28}^{+0.32}}$ & $1.68_{-0.53}^{+0.33}$ \\
    $\alpha$ (deg) & $27.2_{-5.1}^{+3.3}$ & $7.7_{-21.2}^{+6.1}$ & $0.0_{-5.3}^{+5.9}$ & $328.3_{-2.1}^{+2.7}$ & $\mathbf{187.6_{-3.4}^{+3.3}}$ & $176.4_{-5.4}^{+3.9}$ \\
    \hline
    \hline
    \end{tabular}
    \label{tab:2L1S_pie}
    %\tablecomments{All fluxes are on an 18th magnitude scale, e.g., $I_{\rm S} = 18 - 2.5 \log(f_{\rm S})$.}
\end{table*}

\subsection{2L1S Parallax Model}

\begin{figure}[htb] 
    \centering
    \includegraphics[width=\columnwidth]{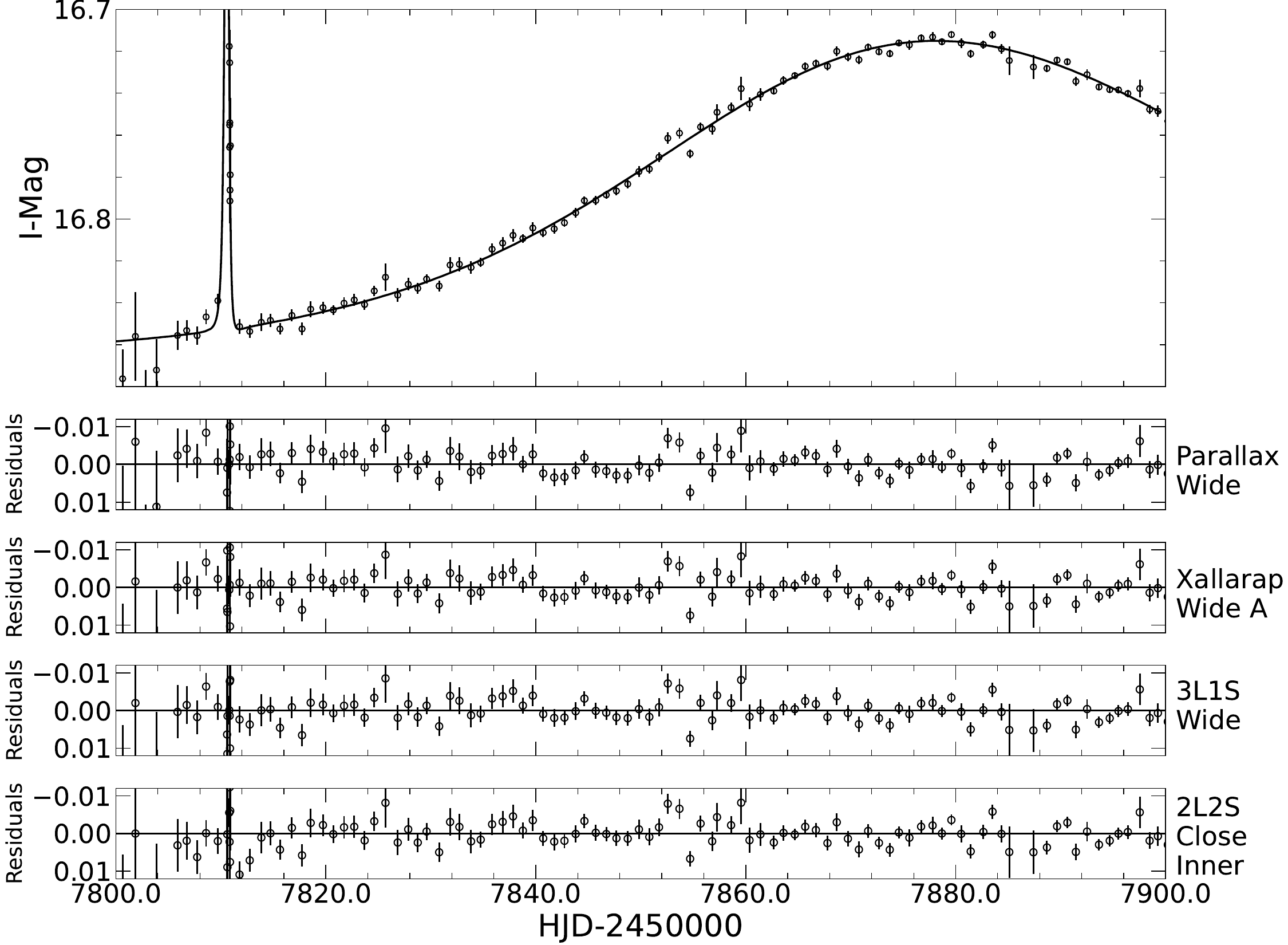}
    \caption{Daily binned data together with the best-fit models of 2L1S parallax, 2L1S xallarap, 3L1S, and 2L2S. All of the four models can remove the long-term residuals shown in Figure \ref{fig:lc2}.}
    %\caption{Observed data together with the best-fit 2L1S parallax solutions for the Close Inner, Close Outer and Wide topology. The Inner and Outer solutions shown in Figures \ref{fig:cau1} and \ref{fig:lc2} of the Wide topology merge into one solution.}
    \label{fig:lc3}
\end{figure}

We first try to improve the fit and remove the long-term residuals with the annual microlens-parallax effect \citep{Gould1992,Gould2000}, in which Earth's acceleration around the Sun introduces nonlinear motion to the lens-source relative motion. We introduce two parameters $\pi_{{\rm E, N}}$ and $\pi_{{\rm E, E}}$, the north and east components of the microlensing parallax vector $\bm{\pi}_{\rm E}$ in equatorial coordinates, 
\begin{equation}\label{equ:pie}
    \bm{\pi}_{\rm E} \equiv \frac{\pi_{\rm rel}}{\thetae} \frac{\bm{\mu}_{\rm rel}}{\mu_{\rm rel}},
\end{equation}
where $\pi_{\rm rel}$ and $\bm{\mu}_{\rm rel}$ are the lens-source relative parallax and proper motion, respectively. We also fit the $u_0 > 0$ and $u_0 < 0$ solutions to account for the ``ecliptic degeneracy'' \citep{Jiang2004, Poindexter2005}. The annual microlens-parallax effect can be degenerate with the lens orbital motion effect \citep{MB09387, OB09020}. Hence, we also introduce this effect. The lens orbital motion is parametrized by $\vec{\gamma} =\left(\frac{ds/dt}{s}, \frac{d\alpha}{dt}\right)$, where $ds/dt$ and $d\alpha/dt$ represent the instantaneous changes in the separation and orientation of the two lens components defined at $\hjd = 7880$. We restrict the MCMC trials to bound systems by calculating the ratio of projected kinetic to potential energy \citep{An2002,OB050071D}:
\begin{equation}
    \beta \equiv \left| \frac{{\rm KE}_{\perp}}{{\rm PE}_{\perp}} \right| = \frac{\kappa M_{\odot} {\rm yr}^2}{8\pi^2}\frac{\pie}{\thetae}\gamma^2\left(\frac{s}{\pie + \pi_{\rm S}/\thetae}\right)^3,
\end{equation}
where $\pi_{\rm S}$ is the source parallax. We adopt $\pi_{\rm S} = 0.12$ mas based on the mean distance to clump giant stars in this direction \citep{Nataf2013}. We reject models with  $\beta \ge 0.8$ for unphysical lens systems.

The resulting parameters are listed in Table \ref{tab:2L1S_pie}. The inner and outer solutions of the Wide topology merge into one solution with the high-order effects, so there are three pairs of solutions. The ``Wide $u_0 > 0$'' solution provides the best fit to the observed data, for which the inclusion of the high-order effects significantly improves the fit by $\Delta\chi^2 = 192$ and removes the long-term residuals (Figure \ref{fig:lc3}). However, this solution has a large parallax value of $1.40\pm 0.14$. Such a large parallax (i.e., $\gtrsim 1$) value is of very low probability though not impossible (e.g., \citealt{OB07224,KB181990}). The ``Close Outer $u_0 > 0$'', ``Close Inner $u_0 < 0$'' and ``Wide $u_0 < 0$'' solutions also have large parallax of $\gtrsim 1$. The ``Close Inner $u_0 > 0$'' and ``Close Outer $u_0 < 0$'' solutions have reasonable parallax values of $0.38^{+0.09}_{-0.05}$ and $0.55^{+0.40}_{-0.18}$, respectively, but they are disfavored by $\Delta\chi^2 = 33.7$ and $15.8$ compared to the ``Wide $u_0 > 0$'' solution, respectively. Therefore, we try the other high-order effect, the microlens-xallarap effect, to see whether a reasonable microlens-xallarap model can fit the long-term residuals.

\subsection{2L1S Xallarap Model}

The long-term asymmetry in the light curve can be caused not only by the motion of the observer around the Sun (the microlens-parallax effect) but also by the inverse effect of motion of the source star in a binary system, called the xallarap effect \citep{Griest1992}. Here we consider the xallarap effect with a circular orbit. This effect introduces five additional parameters that can be defined in various ways \citep[e.g.,][]{MB06074,OB130911}. Below we first introduce a new parameterization of the xallarap effect. Then, we discuss our approach to fitting and its results. 

\subsubsection{Parametrization of the Xallarap Effect}

We define the xallarap orbit using five fitted parameters $(\xi_P, \xi_a, \xi_i, \xi_\Omega, \xi_u)$ and one fixed parameter ($t_{0,\xi}$). The fitted parameters are the usual Keplerian parameters of the orbit:
$\xi_P$ -- the orbital period,
$\xi_a$ -- the semi-major axis relative to $\thetae$, 
$\xi_i$ -- the inclincation, 
$\xi_\Omega$ -- the longitude of the ascending node, and 
$\xi_u$  -- the argument of latitude at the reference epoch $t_{0,\xi}$\footnote{The argument of latitude is the sum of the argument of periapsis and the true anomaly at a given epoch: $u(t) = \nu(t) + \omega$ and for a circular orbit we have $u(t) \equiv \nu(t)$.}. We prefer to use the argument of latitude (instead of, e.g., the time of the periapsis passage) because periapsis is not defined for a circular orbit and is poorly constrained for an eccentric orbit with a small eccentricity. We define the orbital parameters with the reference plane to be the plane of the sky and the reference direction to be the relative lens-source proper motion direction. 

To calculate the influence of the xallarap effect on the relative lens-source position, we calculate the position ($\vec{r_1}(t)$) of the luminous source relative to the center of mass on the reference plane for every epoch  using standard orbit integration. We also calculate this position for the reference epoch: $\vec{r_1}(t_{0,\xi})$. The xallarap shift is calculated as $\vec{r_1}(t) - \vec{r_1}(t_{0,\xi})$. In this approach, the xallarap effect weakly affects the magnification for epochs close to $t_{0,\xi}$. The xallarap shift for the source center of mass is $-\vec{r_1}(t_{0,\xi})$. Furthermore, one can calculate the position of the second source component: the position of the second source relative to the center of mass is $-\vec{r_1}(t)/q_s$ (where $q_s$ is the mass ratio of the source), hence, the xallarap shift for the second source is $-\vec{r_1}(t)/q_s-\vec{r_1}(t_{0,\xi})$.

We note that $t_{0,\xi}$ serves two purposes: it defines the reference epoch for $\xi_u$ and it defines the epoch at which the xallarap does not affect the magnification. The specific choice of $t_{0,\xi}$ value does not have an impact on the former purpose. On the other hand, the specific choice of $t_{0,\xi}$ value for the latter purpose is a very important factor for the convergence of the MCMC chain when one starts the parameter exploration from the best-fit model without xallarap. 

\subsubsection{Xallarap Model Fitting}

We set $t_{0,\xi}$ close to the best-timed part of the event, i.e., anomaly: $t_{0,\xi}\equiv7810.5$. The xallarap introduces five additional parameters and in the case of \event, none of these parameters can be easily estimated from the light curve inspection. We search the parameter space by starting from the static binary-lens models presented in Table~\ref{tab:2L1S}. For each of these four models, we run an MCMC starting with initial positions of the MCMC walkers drawn from uniform distributions $(0, 360)$ of angles $\xi_\Omega$, $\xi_i$, and $\xi_u$. The values of $\xi_a$ were drawn log-uniformly from $(0.001, 0.1)$. For the xallarap period $\xi_P$, we adopt a grid approach. We run the MCMC 10 times independently each one exploring a range of periods: $(3\xi_{P,i}/4, 4\xi_{P,i}/3)$, where $\xi_{P,i}$ is a geometric series of ten elements from 5 to 400 days. We can cover the range of periods from 3.7 to 540 days with overlapping runs, which ensures that the right period should be found even if it is very close to an edge of one of the period ranges. For each run, the initial positions of walkers are drawn from a normal distribution with the mean of $\xi_{P,i}$ and sigma of $0.001$ days. From each of the ten runs, we extract the smallest $\chi^2$ model. Then, we re-run the fitting with starting points randomly drawn very close to these smallest $\chi^2$ models but without limiting the $\xi_P$ values. We then identified and ignored duplicated results and runs which converged to models with much higher $\chi^2$. Because the 2L1S parallax models find that the lens orbital motion effect has almost no influence on the models, the xallarap fitting does not include this effect.

\begin{figure*}
    \centering
    \includegraphics[width = 0.95\columnwidth]{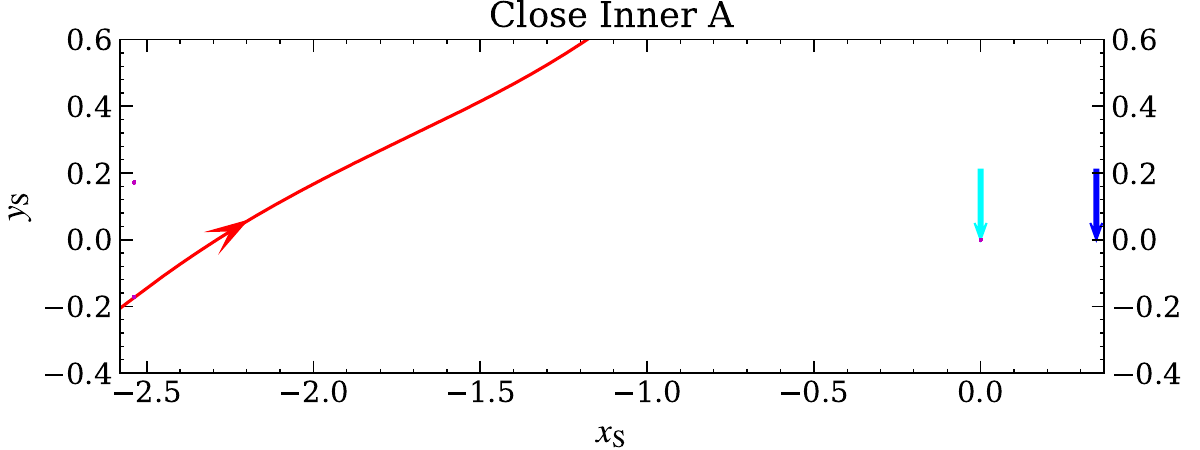}
    \includegraphics[width = 0.95\columnwidth]{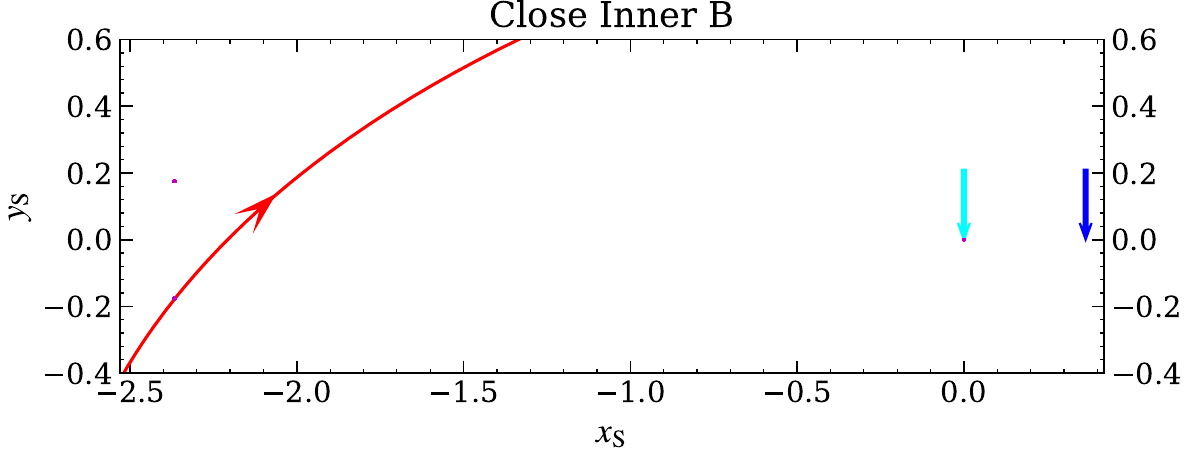}
    \includegraphics[width = 0.95\columnwidth]{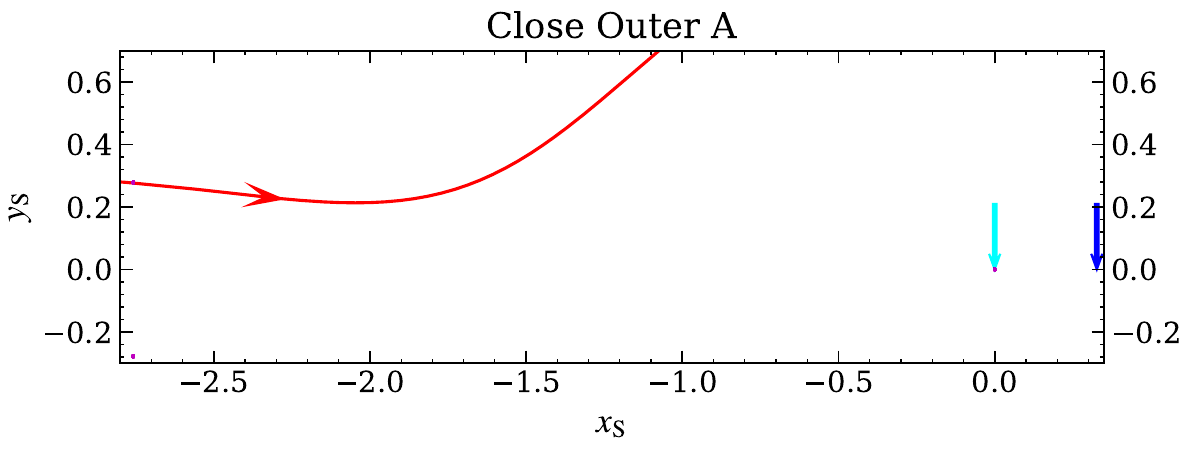}
    \includegraphics[width = 0.95\columnwidth]{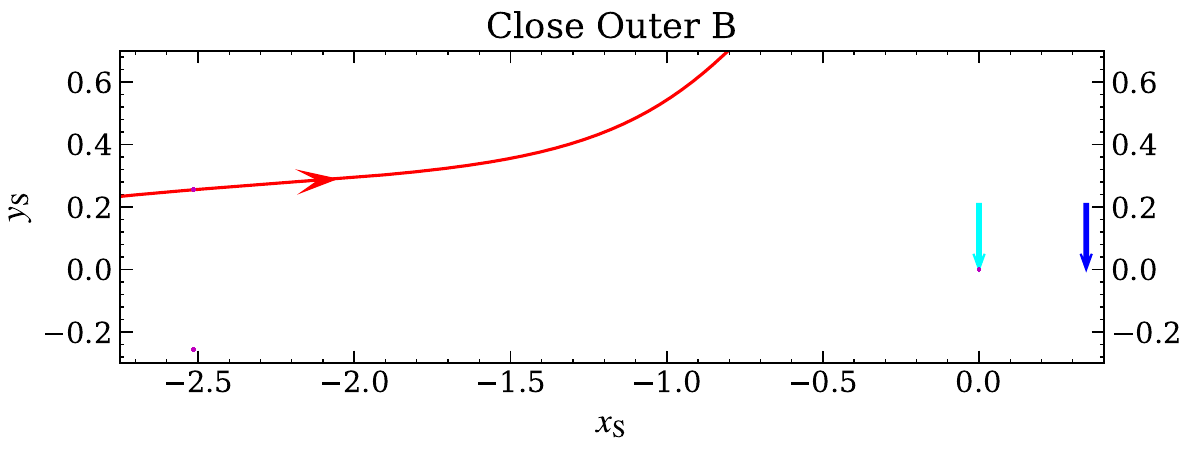}
    \includegraphics[width = 0.95\columnwidth]{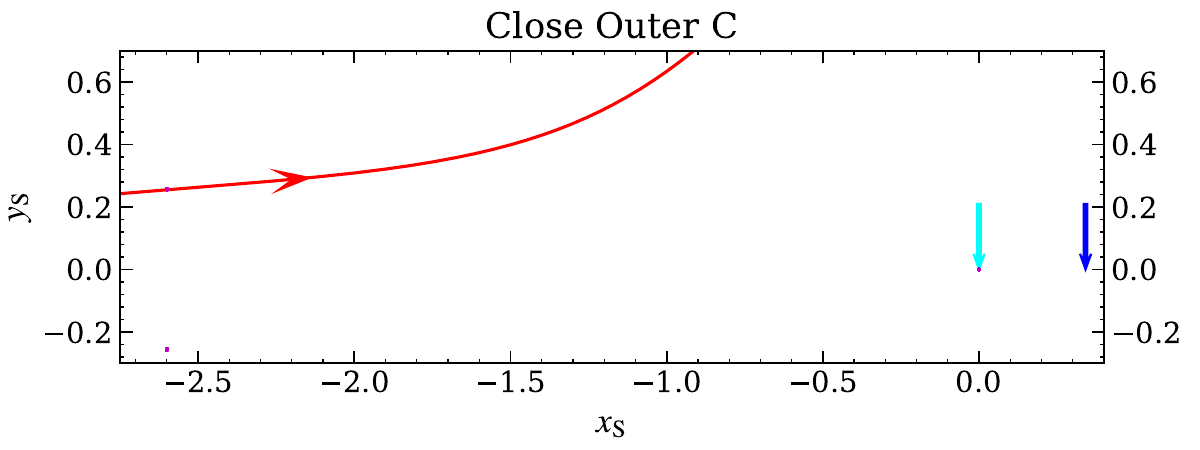}
    \includegraphics[width = 0.95\columnwidth]{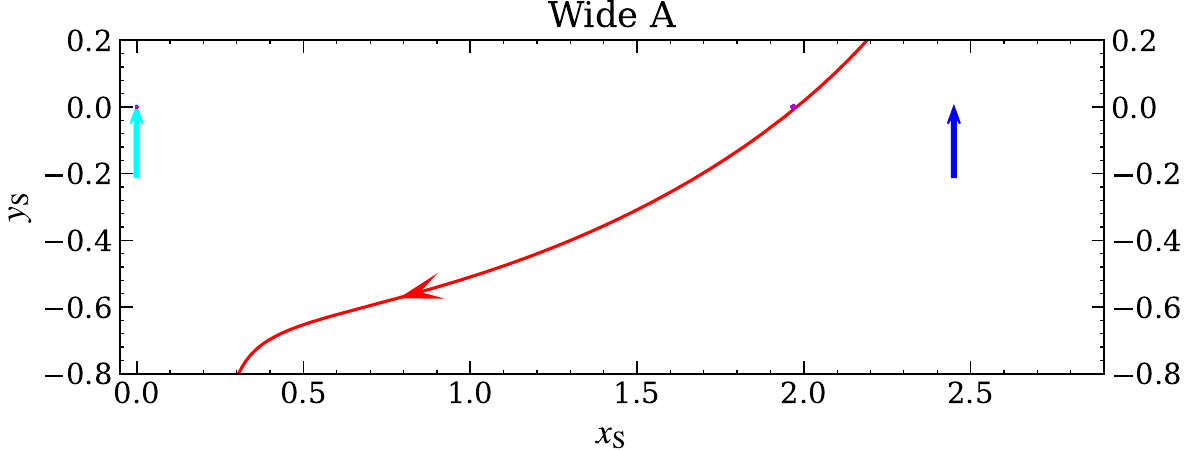}
    \includegraphics[width = 0.95\columnwidth]{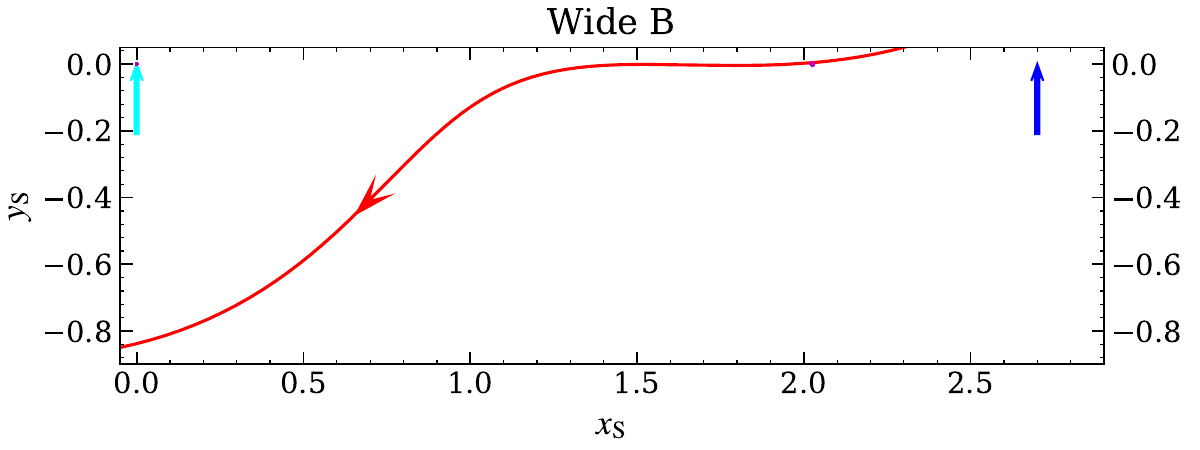}
    \includegraphics[width = 0.95\columnwidth]{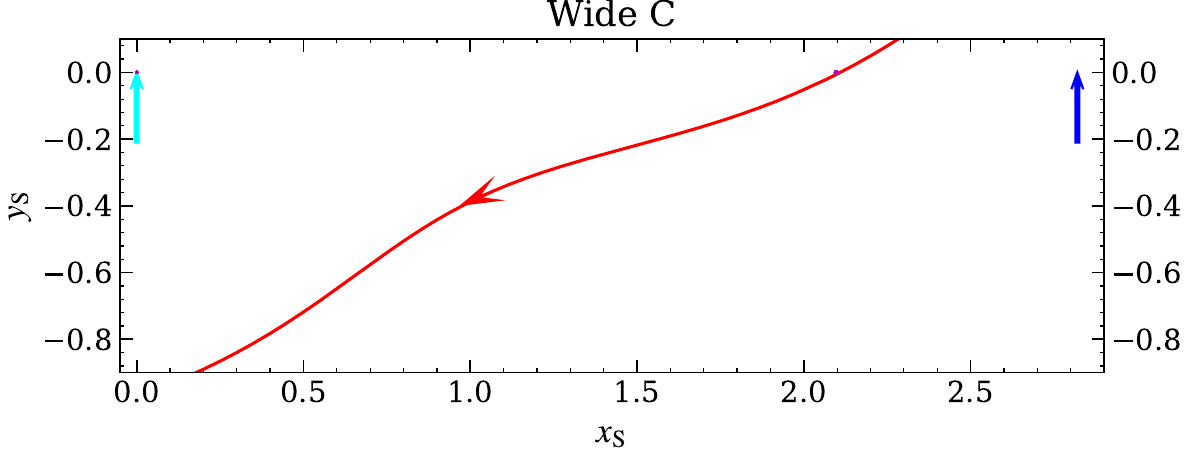}
    \caption{Source trajectories of the 2L1S xallarap models. The symbols are similar to those in Figure \ref{fig:cau1}. Their parameters are given in Tables \ref{tab:2L1S_xa1} and \ref{tab:2L1S_xa2}. The ``Close Outer B'' and ``Close Outer C'' models have the almost same trajectories but their source radii are different.}
    \label{fig:cau2}
\end{figure*}

This model exploration resulted in eight solutions within $\Delta\chi^2 < 30$, including three (labeled as ``A'', ``B'', and ``C'') for the ``Wide'' topology, three (labeled as ``A'', ``B'', and ``C'') for the ``Close Outer'' topology and two (labeled as ``A'' and ``B'') for the ``Close Inner'' topology. Their parameters are presented in Tables \ref{tab:2L1S_xa1} and \ref{tab:2L1S_xa2}, and their source trajectories are shown in Figure \ref{fig:cau2}. All of the eight 2L1S xallarap solutions provide better fits than the best-fit 2L1S parallax solution, with $\Delta\chi^2$ of between 1.1 and 14.3. Therefore, we keep all of the 2L1S xallarap solutions and evaluate all the 2L1S solutions with high-order effects by combining the physical parameters of the lens and the source systems.

%\textbf{\citep{OB09020} - use this convention and claim it.}
%\textbf{$\xi_P$ $\xi_a$ $\xi_i$ $\xi_\Omega$ $\xi_u$ }

\begin{table*}[htb]
    \renewcommand\arraystretch{1.25}
    \centering
    \caption{Lensing Parameters for Close ($s < 1$) Xallarap Models}

    \begin{tabular}{c|c c| c c c}
\hline\hline
& \multicolumn{2}{c|}{Close Inner} & \multicolumn{3}{c}{Close Outer} \\
& A & B & A & B & C \\
\hline
$\chi^2$/dof  & $ 10080.2/10682 $  & $ 10074.7/10682$ & $ 10078.1/10682 $  & $ 10084.6/10682 $  & $ 10084.9/10682 $ \\
\hline
$t_0$   & $ 7882.2\pm0.5 $  & $ 7881.1^{+2.6}_{-7.3} $ & $ 7881.2\pm0.8 $  & $ 7883.0^{+1.5}_{-2.3} $  & $ 7883.7^{+1.0}_{-1.7} $ \\
$u_0$  & $ 1.20^{+0.05}_{-0.08} $  & $ 1.57\pm0.23 $ & $ 1.03\pm0.08 $  & $ 1.21\pm0.16 $  & $ 1.13^{+0.17}_{-0.13} $ \\
$\te$ (days)  & $ 32.0^{+1.5}_{-1.0} $  & $ 39.9\pm2.8 $ & $ 28.0^{+1.5}_{-1.2} $  & $ 32.0\pm2.3 $  & $ 31.3\pm1.9 $ \\
$\rho\,(10^{-2})$  & $ 0.078\pm0.028 $  & $ 0.088\pm0.041 $ & $ 0.044^{+0.047}_{-0.031} $  & $ 0.033^{+0.031}_{-0.024} $  & $ 0.336^{+0.085}_{-0.062} $ \\
$t_{\rm 0,pl}$  & $ 7813.5\pm0.3 $  & $ 7815.9^{+1.1}_{-0.9} $ & $ 7808.2\pm0.3 $  & $ 7807.1\pm0.8 $  & $ 7807.5\pm0.6 $  \\
$u_{\rm 0,pl}$ & $ 0.155^{+0.019}_{-0.016} $  & $ 0.134\pm0.018 $ & $ -0.255^{+0.026}_{-0.037} $  & $ -0.245\pm0.023 $  & $ -0.246\pm0.024 $ \\
$t_{\rm E,pl}$ (days) & $ 1.06\pm0.10 $  & $ 1.48\pm0.15 $ & $ 1.34\pm0.13 $  & $ 1.56^{+0.21}_{-0.15} $  & $ 1.53\pm0.16 $  \\
$\xi_P$ (days) & $ 77^{+16}_{-8} $  & $ 150^{+36}_{-14} $ & $ 101^{+6}_{-5} $  & $ 172^{+21}_{-15} $  & $ 166\pm16 $  \\
$\xi_a$ &  $ 0.049^{+0.013}_{-0.009} $  & $ 0.320^{+0.120}_{-0.070} $ & $ 0.233\pm0.053 $  & $ 0.358^{+0.093}_{-0.077} $  & $ 0.325\pm0.081 $ \\
$\xi_\Omega$ (deg)  & $ 295\pm11 $  & $ 136^{+8}_{-17} $ & $ 96^{+3}_{-3} $  & $ 135\pm6 $  & $ 313\pm7 $  \\
$\xi_i$ (deg) & $ 112^{+10}_{-18} $  & $ 129^{+5}_{-4} $ & $ 117^{+3}_{-4} $  & $ 115\pm8 $  & $ 111^{+7}_{-9} $  \\
$\xi_u$ (deg)  & $ 167^{+41}_{-30} $  & $ 34^{+21}_{-15} $ & $ 77\pm14 $  & $ 111\pm16 $  & $ 291\pm13 $  \\
$I_{\rm S}$ & $18.20^{+0.17}_{-0.18}$ & $17.98^{+0.24}_{-0.22}$  & $18.13^{+0.19}_{-0.22}$ & $17.96^{+0.24}_{-0.22}$ & $18.00^{+0.27}_{-0.25}$ \\
\hline
$s$  & $ 0.347^{+0.014}_{-0.009} $  & $ 0.365^{+0.026}_{-0.019} $ & $ 0.327^{+0.013}_{-0.008} $  & $ 0.343^{+0.018}_{-0.015} $  & $ 0.341\pm0.016 $ \\
$q\,(10^{-4})$  & $ 10.9\pm2.0 $  & $ 13.7^{+2.9}_{-2.3} $  & $ 22.3^{+5.5}_{-3.6} $  & $ 24.1^{+3.5}_{-2.8} $  & $ 23.7\pm3.8 $ \\
$\alpha$ (deg)  & $ 32.15^{+0.57}_{-0.73} $  & $ 46.0^{+6.4}_{-4.1} $ & $ 16.6\pm1.7 $  & $ 22.1\pm3.7 $  & $ 20.3\pm3.1 $  \\
\hline\hline
\end{tabular}
\label{tab:2L1S_xa1}
\end{table*}

\begin{table}[htb]
    \renewcommand\arraystretch{1.25}
    \centering
    \caption{Lensing Parameters for Wide ($s > 1$) Xallarap Models}
    \begin{tabular}{c|c c c}
    \hline
    \hline
    & \multicolumn{3}{c}{Wide} \\
 & A & B & C  \\
\hline
$\chi^2$/dof & $\mathbf{10071.7/10682}$  & $10074.2/10682$  & $ 10077.9/10682$  \\
\hline
$t_0$ & $ \mathbf{7881.63^{+2.30}_{-4.87}} $  & $ 7882.10^{+0.72}_{-0.91} $  & $ 7882.47^{+0.47}_{-0.59} $  \\
$u_0$ & $ \mathbf{1.34\pm0.28} $  & $ 0.92^{+0.25}_{-0.19} $  & $ 1.08^{+0.14}_{-0.21} $  \\
$\te$ (days) & $ \mathbf{45.58\pm8.40}$  & $ 33.45^{+3.98}_{-3.06} $  & $ 32.51^{+3.72}_{-2.23} $  \\
$\rho$ ($10^{-2}$) & $ \mathbf{0.74^{+0.42}_{-0.27}} $  & $ 0.66^{+0.38}_{-0.56} $  & $ 0.60\pm0.43 $  \\
$t_{\rm 0,pl}$ & $ \mathbf{7810.56^{+0.11}_{-0.17}} $  & $ 7810.61^{+0.06}_{-0.11} $  & $ 7810.52^{+0.11}_{-0.09} $  \\
$u_{\rm 0,pl}$ & $ \mathbf{-0.005^{+0.011}_{-0.004}} $  & $ 0.002^{+0.002}_{-0.010} $  & $ 0.001^{+0.004}_{-0.009} $  \\
$t_{\rm E,pl}$ (days) & $ \mathbf{0.45\pm0.13} $  & $ 0.25^{+0.05}_{-0.03} $  & $ 0.25\pm0.03 $  \\
$\xi_P$ (days) & $ \mathbf{155.5^{+22.7}_{-12.0}} $  & $ 93.8\pm10.6 $  & $ 87.0\pm8.8 $  \\
$\xi_a$ & $ \mathbf{0.326^{+0.096}_{-0.060}} $  & $ 0.132\pm0.059 $  & $ 0.085^{+0.040}_{-0.022} $  \\
$\xi_\Omega$ (deg) & $ \mathbf{143.0^{+6.5}_{-9.4}} $  & $ 279.7^{+9.5}_{-4.1} $  & $ 104.0^{+9.0}_{-5.9} $  \\
$\xi_i$ (deg) & $ \mathbf{132.4\pm3.7} $  & $ 112.5^{+4.8}_{-8.3} $  & $ 113.9\pm6.9 $  \\
$\xi_u$ (deg) & $ \mathbf{60.6\pm20.0} $  & $ 236.8^{+22.5}_{-38.4} $  & $ 29.9\pm29.0 $  \\
$I_{\rm S}$ & $\mathbf{17.35^{+0.50}_{-0.18}}$ & $17.56^{+0.52}_{-0.23}$ & $17.78^{+0.44}_{-0.35}$ \\
\hline
$s$ & $ \mathbf{2.45^{+0.40}_{-0.30}} $  & $ 2.70\pm0.25 $  & $ 2.82^{+0.19}_{-0.26} $  \\
$q$ ($10^{-4}$) & $ \mathbf{0.98^{+0.31}_{-0.21}} $  & $ 0.55 \pm 0.14 $  & $ 0.57 \pm 0.11 $  \\
$\alpha$ (deg) & $\mathbf{220.6 \pm 3.9}$ & $203.1^{+3.3}_{-2.6}$ & $206.2^{+1.5}_{-2.1}$  \\
\hline\hline
\end{tabular}
\label{tab:2L1S_xa2}
\end{table}

\section{Color-Magnitude Diagram}\label{sec:CMD}

\begin{figure}[htb] 
    \centering
    \includegraphics[width=\columnwidth]{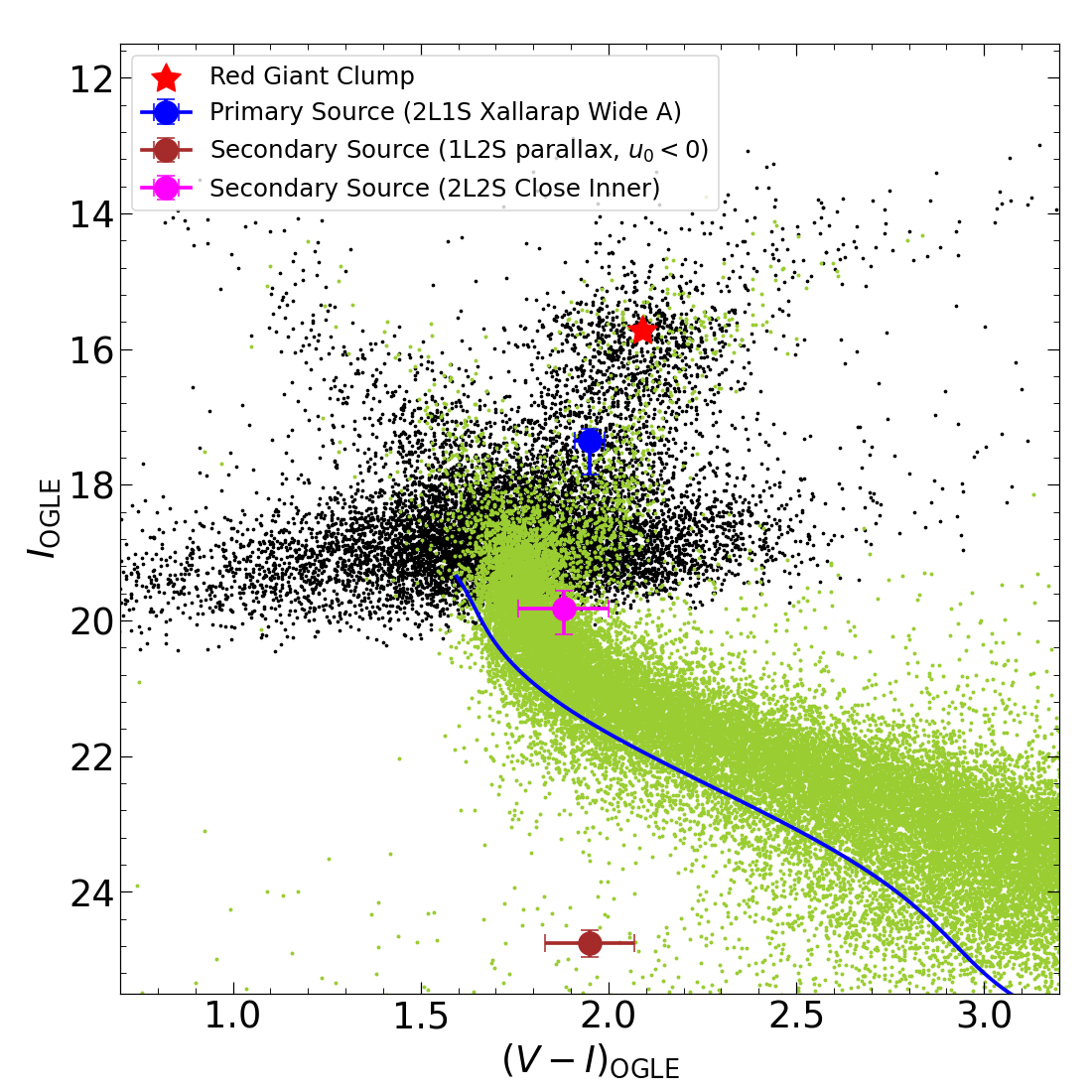}
    \caption{CMD (black points) within $120^{\prime\prime}$ of \event\ using the OGLE-III star catalog \citep{OGLEIII}. The red asterisk indicates the centroid of the red giant clump. The blue line indicates the blue boundary of the bulge main-sequence stars \citep{KB220440}. The yellow-green points shows the HST CMD of \cite{HSTCMD}. The blue dot represents the source position for the 2L1S xallarap ``Wide A'' model. For other models, the source color is the same and the source brightness depends on particular models (Tables \ref{tab:2L1S_xa1} and \ref{tab:2L1S_xa2}). The brown and magenta dots show the secondary sources for the 1L2S parallax ($u_0 < 0$) and 2L2S ``Close Inner'' models.}
    \label{fig:cmd}
\end{figure}

Before the 1L2S analysis, we analyze the CMD to obtain the source color and the source angular radius, $\theta_*$, which are used to exclude the 1L2S model in Section \ref{sec:1L2SpiE} and estimate the lens physical parameters in Section \ref{lens}. We locate the source on a $V - I$ versus $I$ CMD, as shown in Figure \ref{fig:cmd}. The CMD is constructed using the OGLE-III catalog stars \citep{OGLEIII} within $120^{\prime\prime}$ centered on the event. The centroid of the red giant clump is $(V - I, I)_{\rm cl} = (2.09 \pm 0.01, 15.73 \pm 0.01)$. From \cite{Bensby2013} and \cite{Nataf2013}, the intrinsic color and de-reddened magnitude of the red giant clump are $(V - I, I)_{\rm cl,0} = (1.06 \pm 0.03, 14.49 \pm 0.04)$, indicating $A_I = 1.24 \pm 0.04$ and $E(V - I) = 1.03 \pm 0.03$ toward this direction. 

For the source color, we first measure $(V - I)_{\rm S, KMTC} = 2.01 \pm 0.04$ by regression of the KMTC01 $V$ versus $I$ flux and then obtain $(V - I)_{\rm S} = 1.95 \pm 0.04$ by matching the KMTC01 pyDIA CMD and the OGLE-III CMD. Then, the intrinsic source color is $(V - I)_{\rm S, 0} = 0.92 \pm 0.05$. Because the source apparent brightness varies from the models, we first obtain the source angular radius $\theta_*$ for $I_{\rm S} = 17.35$ and then provide a scaling relation for different source brightnesses. Using the color/surface-brightness relation of \cite{Adams2018}, we obtain 
\begin{equation}
    \theta_* = 2.33 \pm 0.12~\mu {\rm as}.
\end{equation}
Here the 5\% error is from Table 3 of \cite{Adams2018}. For any particular model, one can derive $\theta_* = 2.33 \times 10^{-0.2(I_{\rm S} - 17.35)}$. We summarize, $\theta_*$, and the derived $\thetae$ and $\mu_{\rm rel}$ for 2L1S parallax and xallarap models in Table \ref{source}.

\begin{table*}
    \renewcommand\arraystretch{1.5}
    \centering
    \caption{$\theta_*$, $\thetae$ and $\mu_{\rm rel}$ for the 2L1S Models}
    \begin{tabular}{c|c c c c}
    \hline
    \hline
    Model & $I_{\rm S,0}$ & $\theta_*$ ($\mu$as) & $\thetae$ (mas) & $\mu_{\rm rel}$ (${\rm mas\,yr^{-1}}$) \\
    \hline    
    \textbf{2L1S Parallax} & & & & \\
    Close Inner ($u_0 > 0$) & $16.30_{-0.06}^{+0.07}$ & $2.13^{+0.09}_{-0.09}$ & $>0.33$ & $>4.1$ \\
    Close Inner ($u_0 < 0$) & $16.32_{-0.07}^{+0.12}$ & $2.11^{+0.11}_{-0.12}$ & $0.26^{+0.16}_{-0.05}$ & $3.48^{+1.44}_{-0.48}$ \\
    Close Outer ($u_0 > 0$) & $16.43_{-0.14}^{+0.26}$ & $2.00^{+0.15}_{-0.23}$ & $0.29^{+0.04}_{-0.04}$ & $2.59^{+0.62}_{-0.59}$ \\
    Close Outer ($u_0 < 0$) & $16.27_{-0.05}^{+0.05}$ & $2.16^{+0.08}_{-0.08}$ & $>0.44$ & $>5.1$ \\
    Wide ($u_0 > 0$) & $16.49_{-0.16}^{+0.18}$ & $1.95^{+0.15}_{-0.16}$ & $0.11^{+0.02}_{-0.01}$ & $1.28^{+0.23}_{-0.12}$  \\
    Wide ($u_0 < 0$) & $16.30_{-0.06}^{+0.13}$ & $2.13^{+0.10}_{-0.14}$ & $0.15^{+0.16}_{-0.05}$ & $1.90^{+1.84}_{-0.54}$  \\
    \hline
    \textbf{2L1S Xallarap} & & & &  \\
    Close Inner A & $16.96^{+0.17}_{-0.18}$ & $1.57^{+0.15}_{-0.14}$ & $>0.90$ & $>10.4$ \\
    Close Inner B & $16.74^{+0.24}_{-0.22}$ & $1.74^{+0.20}_{-0.20}$ & $>0.67$ & $>6.2$ \\
    Close Outer A & $16.89^{+0.19}_{-0.22}$ & $1.62^{+0.19}_{-0.16}$ &  $>0.46$ & $>6.0$ \\
    Close Outer B & $16.72^{+0.24}_{-0.22}$ & $1.76^{+0.20}_{-0.20}$ & $>1.36$ & $>12.4$ \\
    Close Outer C & $16.76^{+0.27}_{-0.25}$ &  $1.73^{+0.23}_{-0.23}$ & $>0.28$ &  $>3.3$ \\
    Wide A & $16.11^{+0.50}_{-0.18}$ & $2.33^{+0.23}_{-0.50}$ & $>0.09$ & $>0.72$ \\ 
    Wide B & $16.32^{+0.52}_{-0.23}$ & $2.11^{+0.26}_{-0.47}$ & $>0.11$ &  $>1.2$ \\
    Wide C & $16.46^{+0.44}_{-0.35}$ & $1.91^{+0.35}_{-0.36}$ &  $>0.11$ &  $>1.2$ \\
    \hline
    \hline
    \end{tabular}
    \tablecomments{The lower limits for $\thetae$ and $\mu_{\rm rel}$ are at $3\sigma$.}
    \label{source}
\end{table*}

\section{Single-lens Binary-source Model}\label{sec:1L2S}

The total magnification of a 1L2S model is the superposition of the 1L1S magnification of two sources \citep{MB12486}:
\begin{equation}\label{equ:total_mag}
    A_{\lambda} = \frac{A_{1}f_{1,\lambda} + A_{2}f_{2,\lambda}}{f_{1,\lambda} + f_{2,\lambda}} = \frac{A_{1} + q_{f,\lambda}A_{2}}{1 + q_{f,\lambda}}, 
\end{equation}
\begin{equation}\label{equ:bright_ratio}
    q_{f,\lambda} = \frac{f_{2,\lambda}}{f_{1,\lambda}}.
\end{equation}
Here $A_\lambda$ is total magnification at wavelength $\lambda$, and $f_{{\rm i},\lambda}$ is the baseline flux of each source, with $i=1$ and $2$ corresponding to the primary and the secondary sources, respectively. In the following subsections, we consider two cases: a static binary source (including the microlens-parallax effect) and binary source with the xallarap effect.

\subsection{1L2S Parallax Model}\label{sec:1L2SpiE}

Table \ref{tab:1L2S} lists the parameters of the best-fit 1L2S model (derived using MCMC). It is disfavored by $\Delta\chi^2 = 78$ compared to the best-fit 2L1S xallarap model, which is significant enough to rule out the 1L2S parallax model. The model also has a large parallax value ($>1$). Moreover, \cite{Gaudi1998} suggested that the 1L2S model can be excluded by the color difference expected for the two sources with different brightness. For the present case, the 1L2S parallax model indicates almost the same color (i.e., $q_{{f,V}}/q_{{f,I}} \sim 1$) for the two sources with about a 7.5 magnitude difference. According to the CMD analysis in Section \ref{sec:CMD}, the second source has a color of $(V - I) = 1.95 \pm 0.12$. Applying the blue boundary of the bulge main-sequence stars derived by \cite{KB220440}, a source star with $I \sim 24.8$ should be redder than $(V - I) = 2.9$ (see Figure \ref{fig:cmd}). Therefore, the 1L2S parallax model can also be excluded by $\sim 8\sigma$ from the color argument. 

\begin{table}[htb]
    \renewcommand\arraystretch{1.25}
    \centering
    \caption{Lensing Parameters for 1L2S Parallax Models}
    \begin{tabular}{c|c c}
    \hline
    \hline
    Parameters & $u_1 > 0$ & $u_1 < 0$ \\
    \hline
    $\chi^2$/dof & $10188.4/10684$ & $10149.6/10684$ \\
    \hline
    $t_{0,1}$ & $7878.35_{-0.23}^{+0.24}$ & $7877.63_{-0.44}^{+0.39}$ \\
    $u_{0,1}$ & $1.195_{-0.031}^{+0.017}$ & $-1.200_{-0.024}^{+0.053}$ \\
    $\te$ (days) & $45.2_{-1.4}^{+1.2}$ & $53.6_{-3.6}^{+3.4}$ \\
    $t_{0,2}$ & $7818.8_{-1.8}^{+1.4}$ & $7824.8_{-6.0}^{+4.8}$ \\
    $u_{0,2}$ & $-0.877_{-0.028}^{+0.041}$ & $1.252_{-0.090}^{+0.067}$ \\
    $\rho_2$ ($10^{-2}$) & $2.43_{-0.19}^{+0.24}$ & $3.34_{-0.52}^{+0.50}$ \\
    $q_{{f,I}}$ ($10^{-3}$) & $0.76_{-0.06}^{+0.07}$ & $1.08_{-0.12}^{+0.15}$ \\
    $q_{{f,V}}$ ($10^{-3}$) & $0.76_{-0.09}^{+0.11}$ & $1.07_{-0.15}^{+0.17}$ \\
    $\pi_{\rm E, N}$ & $-1.389_{-0.042}^{+0.064}$ & $2.042_{-0.156}^{+0.114}$ \\
    $\pi_{\rm E, E}$ & $-0.388_{-0.048}^{+0.064}$ & $-0.282_{-0.115}^{+0.155}$ \\
    $I_{\rm S, total}$ & $17.30_{-0.03}^{+0.06}$ & $17.33_{-0.05}^{+0.11}$ \\
    \hline
    $q_{{f,V}}/q_{{f,I}}$ & $1.00_{-0.10}^{+0.12}$ & $0.99_{-0.11}^{+0.11}$ \\
    \hline
    \hline
    \end{tabular}
    \tablecomments{$I_{\rm S, total}$ is derived from the total fluxes of the two sources.}
    \label{tab:1L2S}
\end{table}

\subsection{1L2S Xallarap Model}

We consider a model with a single lens, two luminous sources, and full Keplerian motion of the source components (i.e., xallarap). In this model, the anomaly is produced by the approach of the secondary source to the lens. Such a model has a low {\it a priori} probability because the xallarap effect has to produce a significant difference in the observed timescales of the two subevents.

For the MCMC fitting of such a model, one has to start with a model that produces the anomaly at the right time. If one sets to xallarap parameters to some random values and starts MCMC from them, then the exploration of the parameter space will be extremely inefficient: most sets of xallarap parameters will not produce anomaly at the right time. In order to find starting parameters for MCMC, we performed several steps described below.

First, we randomly draw many sets of parameters. We consider two types of xallarap orbits: circular and eccentric. For circular orbits we use normal distributions for ($t_0$, $u_0$, $\te$) with means found in a fit without the anomaly and small dispersions. Then, $\xi_P$ is drawn from a log-uniform from 50 to 500 days, $\xi_a$ is drawn from a log-uniform distribution from 0.01 to 0.5, angles ($\xi_i$, $\xi_\Omega$, $\xi_u$) are from uniform distributions, and $q_s$ is log-uniform from 0.001 to 0.5. We fix $t_{0,\xi}$ at $\hjd = 7880$ (note that this differs from 2L1S choice). For eccentric orbits, there are two additional parameters: eccentricity of xallarap ($\xi_e$) drawn from log-uniform distribution from 0.01 to 1 and argument of periapsis ($\xi_\omega$) drawn from a uniform distribution. After drawing $2\times10^{8}$ circular and the same number of eccentric orbits, we select the ones for which the distance between the second source and the lens at $\hjd = 7810.5$ was smaller than 0.1, because the second source is much fainter and thus needs to pass very close to the lens to produce a significant anomaly. There were around 26,000 such orbits of each type. Then we calculated $\chi^2$ values for these models without any constraints on fluxes. We selected 100 models with the smallest $\chi^2$ in each case and for them we run the MCMC with only two parameters fitted: the ratio of the source fluxes for the $I$ band, $q_{{f,I}}$, and the ratio for the $V$ band, $q_{{f,V}}$. We further narrow down the models considered to the ones with $\chi^2$ below a threshold value of 10350, which resulted in 20 circular orbits and 24 eccentric orbits.

The above multi-step procedure allows us to go from $4\times10^8$ randomly drawn sets of parameters to 44 that have light curves resembling the observed one. Importantly, the above procedure is robust and efficient. For each of the above models we then run MCMC with the fitting of all the microlensing parameters: three PSPL ones, five basic xallarap ones (plus two more for eccentric orbits), the mass ratio of the two sources, $q_{\rm S}$, and two ratios of source fluxes. In total, there are 11 parameters for circular orbits and 13 for eccentric ones. These MCMC runs include the constraint on the maximum source flux as other models described before (this constraint is not used in calculations presented in the previous paragraph). We found the smallest $\chi^2$ of 10082.1. However, this model is unphysical: $q_{\rm S} = 0.00399 \pm 0.00063$ points to a planet-to-star mass ratio but the flux ratio $q_{f,I} = 0.094 \pm 0.015$ points to a G or K dwarf. The other solutions all have planetary or brown-dwarf mass ratios (i.e., $q_{\rm S} < 0.03$) and stellar flux ratios ($q_{f,I} > 10^{-3}$). 

%$q_{f,I}$ from $10^{-3}$ to 0.04.

Some of the models considered have a very close approach of the second source to the lens. Hence, we run another 44 MCMC runs with finite source effects for the second source. We introduce an additional parameter: $\rho_2$ -- the angular size of the second source scaled to $\thetae$. This increases the number of parameters to 12 (circular orbits) or 14 (eccentric orbits). We assume a uniform brightness profile and used single-lens finite-source calculations that implement the \cite{1994ApJ...421L..71G} method. The lowest $\chi^2$ of these runs is 10062.6, however, all source mass ratios are still in the planetary range while flux ratios are in the stellar range. Moreover, the color of the secondary source is also inconsistent with the color of bulge main-sequence stars. We conclude that none of the 1L2S with xallarap models that fit the data well are physical.

%Additionally, $\rho_2$ was significant, which contradicts the mass ratio values and in some cases smaller than the periapsis distance. 

\section{Physical Parameters and Model Preference}\label{lens}
\begin{table*}
    \renewcommand\arraystretch{1.5}
    \centering
    \caption{Physical Parameters from a Bayesian Analysis and the  $\Delta\chi^2$ from different weights}
    \begin{tabular}{c|c c c c c c c|c c c}
    \hline
    \hline
     & \multicolumn{7}{c|}{Physical Parameters} & \multicolumn{3}{c}{$\Delta\chi^2$} \\
     
    & $M_1$ & $M_{\rm planet}$ & $D_{\rm L}$ & $a_{\bot, \rm planet}$ & $\mu_{\rm rel}$ & $M_3$ & $a_{\bot,3}$ & Gal.Mod. & Light Curve & Total \\

    Units & $M_{\odot}$ & $M_{\oplus}$ & kpc & au & ${\rm mas\,yr^{-1}}$ & $M_{\odot}$ & au & & & \\

    \hline
    \textbf{2L1S Parallax} & & & & & & & & \\ 
    Close Inner ($u_0 > 0$) & $0.15^{+0.06}_{-0.05}$ & $86.1^{+68.0}_{-49.6}$ & $3.69^{+0.74}_{-0.84}$ & $0.57^{+0.11}_{-0.11}$ & $5.6^{+1.2}_{-0.8}$ & ... & ... & 6.2 & 30.1 & 34.8 \\
    Close Inner ($u_0 < 0$) & $0.065^{+0.024}_{-0.021}$ & $236^{+211}_{-131}$ & $3.94^{+0.62}_{-1.17}$ & $0.39^{+0.07}_{-0.09}$ & $4.1^{+1.6}_{-0.7}$ & ... & ... & 15.7 & 45.0 & 59.2 \\ 
    Close Outer ($u_0 > 0$) & $0.021^{+0.004}_{-0.002}$ & $69.8^{+5.9}_{-5.5}$ & $1.94^{+0.07}_{-0.44}$ & $0.19^{+0.01}_{-0.02}$ & $4.6^{+1.2}_{-0.2}$ & ... & ... & 19.4 & 36.1 & 54.0 \\
    Close Outer ($u_0 < 0$) & $0.19^{+0.04}_{-0.04}$ & $93.3^{+30.4}_{-20.8}$ & $3.39^{+0.46}_{-0.42}$ & $0.66^{+0.07}_{-0.07}$ & $6.6^{+1.1}_{-0.7}$ & ... & ... & 6.5 & 48.0 & 53.0 \\
    Wide ($u_0 > 0$) & $0.015^{+0.008}_{-0.002}$ & $0.84^{+0.28}_{-0.22}$ & $3.06^{+0.40}_{-0.80}$ & $1.53^{+0.11}_{-0.23}$ & $2.7^{+1.3}_{-0.6}$ & ... & ... & 23.3 & 14.3 & 36.1 \\
    Wide ($u_0 < 0$) & $0.048^{+0.034}_{-0.021}$ & $2.56^{+2.07}_{-1.11}$ & $4.92^{+2.04}_{-1.60}$ & $2.72^{+0.96}_{-0.76}$ & $2.7^{+2.0}_{-0.9}$ & ... & ... & 16.0 & 30.7 & 45.2 \\
    \hline
    \textbf{2L1S Xallarap} & & & & & & & & \\ 
    Close Inner A & $0.52^{+0.44}_{-0.27}$ & $190^{+155}_{-101}$ & $2.25^{+1.89}_{-1.53}$ & $0.87^{+0.52}_{-0.44}$ & $12.8^{+4.5}_{-2.8}$ & ... & ... & 8.3 & 8.5 & 15.3 \\
    Close Inner B & $0.66^{+0.39}_{-0.32}$ & $294^{+187}_{-144}$ & $3.86^{+1.66}_{-2.11}$ & $1.19^{+0.37}_{-0.48}$ & $7.9^{+3.9}_{-1.6}$ & ... & ... & 5.8 & 3.0 & 7.3 \\
    Close Outer A & $0.66^{+0.37}_{-0.34}$ & $476^{+297}_{-248}$ & $6.46^{+1.25}_{-2.36}$ & $1.03^{+0.22}_{-0.28}$ & $6.6^{+2.0}_{-1.1}$ & ... & ... & 2.8 & 6.4 & 7.7 \\
    Close Outer B & $0.41^{+0.27}_{-0.18}$ & $330^{+228}_{-144}$ & $0.68^{+0.61}_{-0.28}$ & $0.49^{+0.27}_{-0.18}$ & $18.8^{+9.0}_{-4.8}$ & ... & ... & 10.6 & 12.9 & 22.0 \\
    Close Outer C & $0.65^{+0.36}_{-0.32}$ & $504^{+304}_{-255}$ & $6.91^{+1.03}_{-2.08}$ & $1.03^{+0.25}_{-0.26}$ & $5.5^{+1.5}_{-1.2}$ & ... & ... & 0.0 & 13.2 & 11.7 \\
    Wide A & $0.50^{+0.38}_{-0.29}$ & $15.6^{+13.8}_{-9.3}$ & $7.53^{+0.87}_{-1.54}$ & $5.75^{+2.39}_{-2.00}$ & $3.2^{+1.6}_{-1.1}$ & ... & ... & 1.5 & 0.0 & 0.0 \\
    Wide B & $0.43^{+0.37}_{-0.25}$ & $6.98^{+8.60}_{-4.61}$ & $7.66^{+0.82}_{-1.26}$ & $5.55^{+2.31}_{-1.73}$ & $3.2^{+1.5}_{-1.0}$ & ... & ... & 0.7 & 2.5 & 1.7 \\
    Wide C & $0.45^{+0.37}_{-0.26}$ & $8.35^{+7.36}_{-4.92}$ & $7.62^{+0.83}_{-1.36}$ & $6.11^{+2.32}_{-2.00}$ & $3.5^{+1.4}_{-1.1}$ & ... & ... & 0.2 & 6.2 & 4.9 \\ 
    \hline
    \textbf{3L1S} & & & & & & & & \\
    Close Inner & $0.24^{+0.11}_{-0.09}$ & $171^{+92}_{-64}$ & $0.45^{+0.21}_{-0.16}$ & $0.40^{+0.13}_{-0.14}$ & $27.9^{+8.5}_{-7.0}$ & $0.21^{+0.09}_{-0.08}$ & $0.29^{+0.09}_{-0.10}$ & 12.3 & 11.3 & 22.1 \\
    Close Outer & $0.68^{+0.37}_{-0.35}$ & $783^{+458}_{-401}$ & $6.22^{+1.17}_{-1.79}$ & $1.50^{+0.29}_{-0.40}$ & $8.7^{+1.2}_{-1.0}$ & $0.48^{+0.27}_{-0.24}$ & $1.04^{+0.20}_{-0.28}$ & 0.8 & 50.7 & 50.0 \\
    Wide & $0.32^{+0.42}_{-0.21}$ & $5.20^{+6.86}_{-3.39}$ & $7.69^{+0.83}_{-1.29}$ & $4.65^{+3.50}_{-1.59}$ & $3.3^{+2.9}_{-1.3}$ & $0.23^{+0.31}_{-0.15}$ & $0.54^{+0.41}_{-0.19}$ & 1.8 & 5.9 & 6.2 \\
    \hline
    \textbf{2L2S} & & & & & & & & \\
    Close Inner & $0.67^{+0.40}_{-0.33}$ & $299^{+186}_{-150}$ & $5.06^{+1.47}_{-1.55}$ & $1.15^{+0.29}_{-0.34}$ & $8.1^{+0.9}_{-0.7}$ & ... & ... & 3.4 & 18.2 & 20.1 \\
    Close Outer & $0.69^{+0.39}_{-0.35}$ & $707^{+426}_{-366}$ & $5.58^{+1.43}_{-1.70}$ & $1.08^{+0.25}_{-0.32}$ & $7.1^{+0.8}_{-0.7}$ & ... & ... & 3.1 & 45.0 & 46.6 \\
    Wide & $0.32^{+0.35}_{-0.18}$ & $7.28^{+8.35}_{-4.12}$ & $7.79^{+0.81}_{-1.08}$ & $5.53^{+1.69}_{-1.19}$ & $2.6^{+0.8}_{-0.5}$ & ... & ... & 2.0 & 21.3 & 21.8 \\
    \hline
    \hline
    \end{tabular}
    \tablecomments{ 
Gal.Mod. represents the relative probability from the Galactic model, for which the $\Delta\chi^2$ is derived by $-2\ln({\rm Gal.Mod.})$. The $\Delta\chi^2$ of light-curve analysis are from Tables \ref{tab:2L1S_pie}, \ref{tab:2L1S_xa1}, \ref{tab:2L1S_xa2}, \ref{tab:3L1S}, and \ref{tab:2L2S}.}
\label{tab:phy}
\end{table*}

From the 2L1S analysis, we obtain six parallax models and eight xallarap models. In this section, we estimate their lenses physical parameters by conducting a Bayesian analysis based on the Galactic model, which also can be used to infer the preferred model (e.g., OGLE-2017-BLG-1806, \citealt{-4planet}). The Galactic model we adopt is the same as the model used by \cite{Yang2021_GalacticModel}. We note that {\it Gaia} \citep{Gaia2016AA, Gaia2018AA} reported a proper motion measurement for the object at the event position. However, we do not adopt this measurement in the Bayesian analysis because {\it Gaia} did not resolve the nearby field star that is $0.5^{\prime\prime}$ away from the source, the event could contain blend, and the {\it Gaia} goodness-of-fit parameter RUWE has high value: 1.65.  

We create a sample of $10^8$ simulated events. For each simulated event $i$ of solution $k$, we weight it by
\begin{equation}
    w_{{\rm Gal},i,k} = \Gamma_{i,k} \times p_{i,k}(\te) p_{i,k}(\thetae) p_{i,k}(\bm{\pi}_{\rm E}),
\end{equation}
where $\Gamma_{i,k} = \theta_{{\rm E},i,k} \times \mu_{{\rm rel},i,k}$ is the microlensing event rate, $\times p_{i,k}(\te) p_{i,k}(\thetae)$ and $p_{i,k}(\bm{\pi}_{\rm E})$ are the likelihood distributions of $t_{\rm E,i,k}$, $\theta_{{\rm E},i,k}$ and $\bm{\pi}_{{\rm E},i,k}$ from light-curve and CMD analysis.  

Table \ref{tab:phy} shows the resulting posterior distributions of the host mass, $M_1$, the planet mass, $M_{\rm planet}$, the lens distance, $D_{\rm L}$, the projected planet-host separation, $a_{\bot, \rm planet}$, and the lens-source relative proper motion, $\mu_{\rm rel}$. Table \ref{tab:phy} also presents the $\Delta\chi^2$ between different models based on the relative probability from the Galactic model and the light-curve analysis. The 2L1S xallarap model ``Wide A'' has the highest probability, and the six 2L1S parallax models are disfavored by $\Delta\chi^2 \geq 34.8$. For four of the 2L1S parallax models, ``Close Inner ($u_0 < 0$)'', ``Close Outer ($u_0 > 0$)'', ``Wide ($u_0 > 0$)'', and ``Wide ($u_0 < 0$)'', the host star is a median- or low-mass brown dwarf located in the Galactic disk and thus the lens number density and Galactic-model likelihood are low. For the other two 2L1S parallax models, ``Close Inner ($u_0 > 0$)'' and ``Close Outer ($u_0 < 0$)'', although their Galactic-model likelihoods are slightly low, they are disfavored by $\Delta\chi^2 \geq 30.1$ from the light-curve analysis. Therefore, we only adopt the xallarap models as our surviving models\footnote{The xallarap models have only one more parameter than the parallax models, so even if we conduct the model selection using Akaike's Information Criterion (AIC) or the Bayesian Information Criterion (BIC), the parallax models are still disfavored by $>5\sigma$ (i.e., $\Delta\chi^2 > 25$ ).}.

\begin{table}[htb]
    \renewcommand\arraystretch{1.25}
    \centering
    \caption{Source companions for the 2L1S xallarap models}
    \begin{tabular}{c|c c c}
    \hline
    \hline
     & $M_{\rm comp} (M_\odot)$ & $M_{\rm comp, limit} (M_\odot)$ & $a_{\rm tot}$ (au) \\
     \hline
     Close Inner A & $13.6^{+28.5}_{-9.1}$ & $>1.0$ & $0.86^{+0.36}_{-0.22}$ \\
     Close Inner B & $1284^{+4012}_{-1012}$ & $>3.7$ & $5.65^{+3.37}_{-2.28}$ \\
     Close Outer A & $1018^{+2805}_{-797}$ & $>4.6$ & $4.28^{+2.37}_{-1.71}$ \\
     Close Outer B & $2063^{+4037}_{-1496}$ & $>33$ & $7.70^{+3.26}_{-2.69}$ \\
     Close Outer C & $4.8^{+5.7}_{-2.6}$ & $>0.39$ & $1.62^{+0.41}_{-0.30}$ \\
     Wide A & $4.4^{+15.6}_{-3.1}$ & $>0.28$ & $0.99^{+0.57}_{-0.24}$ \\
     Wide B & $1.2^{+4.7}_{-0.8}$ & $>0.01$ & $0.53^{+0.23}_{-0.08}$ \\
     Wide C & $0.8^{+3.6}_{-0.5}$ & $>0.02$ & $0.48^{+0.20}_{-0.06}$ \\
    \hline
    \hline
    \end{tabular}
    \tablecomments{$M_{\rm comp, limit}$ is the 3-$\sigma$ lower limit for $M_{\rm comp}$.}
\label{tab:com}
\end{table}

For the eight 2L1S xallarap models, all are within $5\sigma$ and probably have M- or K-dwarf hosts. We further check whether their source systems are physically reasonable. We calculate the source 
semi-major axis by 
\begin{equation}
    a_{\rm S} = \xi_a \thetae D_{\rm S},
\end{equation}
where $D_{\rm S}$ is the source distance and we use its distribution from the Bayesian analysis above. Then, we derive the mass and separation for the source companion by Kepler's third law. Table \ref{tab:com} lists the information for the source companion, including the mass, $M_{\rm com}$, the 3-$\sigma$ lower limit for $M_{\rm com}$, the separation from the source, $a_{\rm tot}$. Three models, ``Close Inner B'', ``Close Outer A'', and ``Close Outer B'', probably require an intermediate-mass black hole source companion, so we exclude them. For the remaining five models, three have a wide topology with projected planet-host separations of $\sim 6$ au and planetary masses between super-Earth-mass and Neptune-mass, two have a close topology with projected planet-host separations of $\sim 1$ au and the planetary masses of $\sim 1$ Jovian mass. The relative lens-source proper motion of the ``Close Inner A'' model is significantly higher than the other four solutions, with $\mu_{\rm rel} \sim 13~{\rm mas\,yr^{-1}}$. The brightness contrast for the source and the lens is $\sim 100$ in the near-infrared band, so it probably requires a separation of $\sim 100$ mas to resolve them by the current high-angular resolution instruments. Therefore, the ``Close Inner A'' model may be tested in 2025 or earlier. The other four solutions, ``Wide A'', ``Wide B'', ``Wide C'', ``Close Outer C'' cannot be distinguished by high-angular resolution imaging due to the similar $\mu_{\rm rel}$.

%Of them, five have $\Delta\chi^2 < 10$ compared to the model ``A'', including three with a wide ($s > 1$) topology and two with a close ($s < 1$) topology. For the wide cases, the projected planet-host separation is $\sim 6$ au, and the planetary mass is between super-Earth-mass and Neptune-mass. For the close cases, the projected planet-host separation is $\sim 1$ au, and the planetary mass is about one Jovian mass. The proper motions are $\sim 3~{\rm mas\,yr^{-1}}$ and $\sim 7~{\rm mas\,yr^{-1}}$ for the wide and close topology, respectively. Therefore, the close/wide degeneracy for the present case may be distinguished with future high-resolution observations. The brightness contrast for the source and the lens is about 100 times in the near-infrared band, so it probably requires adaptive optics (AO) imaging on 30m-class telescopes if the wide cases are true.

\section{Discussion: Four-body Models}\label{dis}

We have followed the ``standard'' light-curve analysis for a microlensing planetary event. That is, we have fitted the observed data with three-body models (2L1S and 1L2S) and tried high-order effects (the parallax, lens orbital motion, and xallarap effects) to fit out the long-term residuals from the static models. We have found that the 2L1S xallarap models fit the observed data well (Figure \ref{fig:lc3}) and the resulting lens physical parameters based on the Bayesian analysis are physically reasonable (Table \ref{tab:phy}). However, we are still wondering whether the long-term residuals can be fitted by adding a fourth body instead of high-order effects. Therefore, we pursue four-body models in Sections \ref{sec:3L1S} and \ref{sec:2L2S} and discuss the implications in Section \ref{sec:imp}.

\subsection{Triple-Lens Single-Source Model}\label{sec:3L1S}

\begin{table}[htb]
    \renewcommand\arraystretch{1.25}
    \centering
    \caption{Lensing Parameters for the 3L1S Models}
    \begin{tabular}{c|c c c c}
    \hline
    \hline
    Parameters & Close Inner & Close Outer & Wide \\
    \hline
    $\chi^2$/dof & $10083.0/10684$ & $10122.4/10684$ & $\mathbf{10077.6/10684}$ \\
    \hline
    $t_0$ & $7884.59_{-0.15}^{+0.17}$ & $7884.31_{-0.32}^{+0.27}$ & $\mathbf{7884.47_{-0.38}^{+0.37}}$ \\
    $u_0$ & $1.288_{-0.018}^{+0.012}$ & $1.246_{-0.051}^{+0.036}$ & $\mathbf{1.276_{-0.052}^{+0.068}}$ \\
    $t_{\rm E}$ (days) & $31.90_{-0.24}^{+0.29}$ & $32.79_{-0.64}^{+0.93}$ & $\mathbf{32.22_{-0.80}^{+1.03}}$ \\
    $\rho$ ($10^{-2}$) & $0.033_{-0.009}^{+0.014}$ & $0.316_{-0.030}^{+0.027}$ & $\mathbf{1.159_{-0.078}^{+0.049}}$ \\
    $s_2$ & $0.4530_{-0.0038}^{+0.0051}$ & $0.4943_{-0.0071}^{+0.0102}$ & $\mathbf{2.8838_{-0.0856}^{+0.0729}}$ \\
    $q_2$ ($10^{-4}$) & $25.26_{-3.59}^{+3.54}$ & $32.83_{-5.13}^{+5.36}$ & $\mathbf{0.49_{-0.05}^{+0.07}}$ \\
    $\alpha$ (deg) & $-25.39_{-0.58}^{+0.44}$ & $-16.65_{-3.39}^{+3.26}$ & $\mathbf{-22.22_{-4.13}^{+4.18}}$ \\
    $s_3$ & $0.3335_{-0.0043}^{+0.0049}$ & $0.3454_{-0.0062}^{+0.0077}$ & $\mathbf{0.3364_{-0.0093}^{+0.0077}}$ \\
    $q_3$ & $0.809_{-0.034}^{+0.038}$ & $0.551_{-0.080}^{+0.092}$ & $\mathbf{0.727_{-0.094}^{+0.144}}$ \\
    $\psi$ (deg) & $-29.02_{-0.62}^{+0.54}$ & $-18.34_{-2.02}^{+1.80}$ & $\mathbf{126.40_{-4.30}^{+3.97}}$ \\
    $I_{\rm S}$ & $17.270_{-0.023}^{+0.035}$ & $17.361_{-0.070}^{+0.100}$ & $\mathbf{17.294_{-0.126}^{+0.109}}$ \\
    \hline
    \hline
    \end{tabular}
\label{tab:3L1S}
\end{table}

First, we add an additional lens component to the static 2L1S model to fit the long-term residuals, i.e., the triple-lens single-source (3L1S) model. Relative to the static 2L1S model, the 3L1S models have three additional parameters, ($s_3, q_3, \psi$), to describe the third body, $M_3$. These are the $M_1$-$M_3$ separation scaled to $\thetae$, the mass ratio of $M_3$ relative to $M_1$, and the orientation angle of $M_3$ with respect to the $M_1$-$M_2$ axis as seen from $M_1$. To avoid confusion, for the 3L1S analysis we designate $s_2$ and $q_2$ for the separation and mass ratio of $M_2$ to $M_1$, respectively.

\begin{figure}[ht] 
    \centering
    \includegraphics[width=.95\columnwidth]{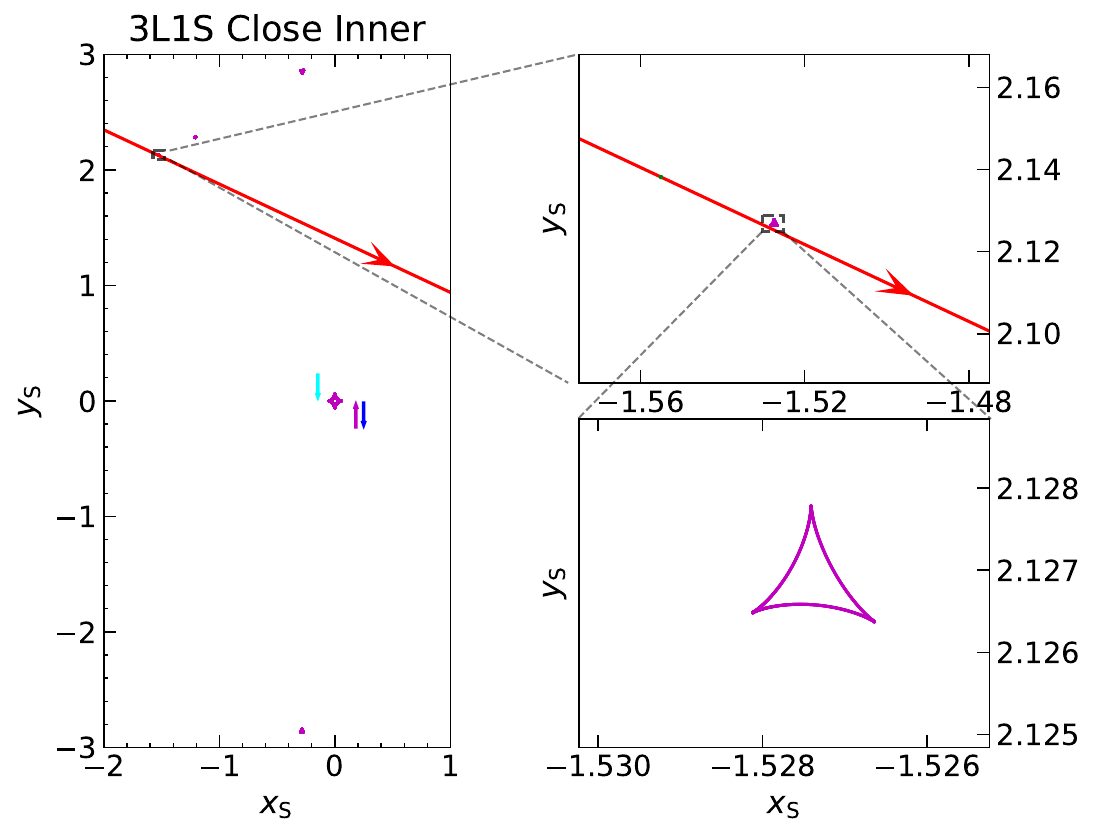}
    \centering
    \includegraphics[width=.95\columnwidth]
    {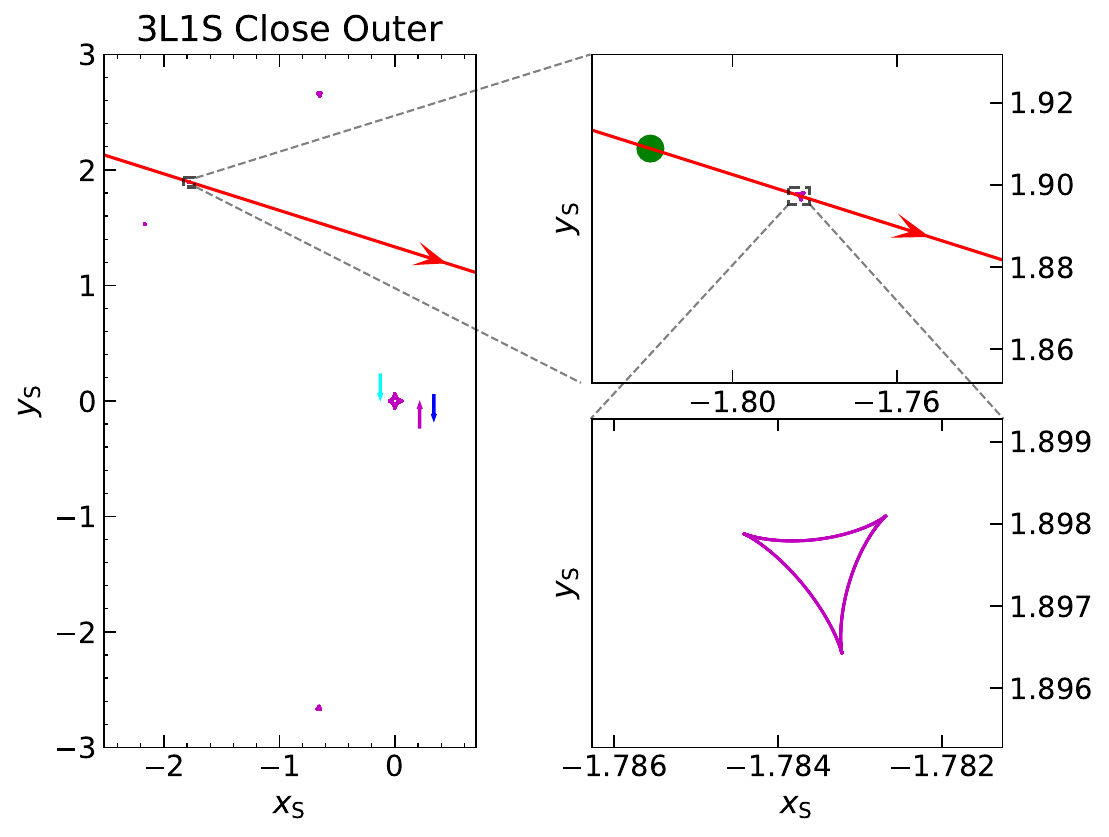}
    \centering
    \includegraphics[width=.95\columnwidth]
    {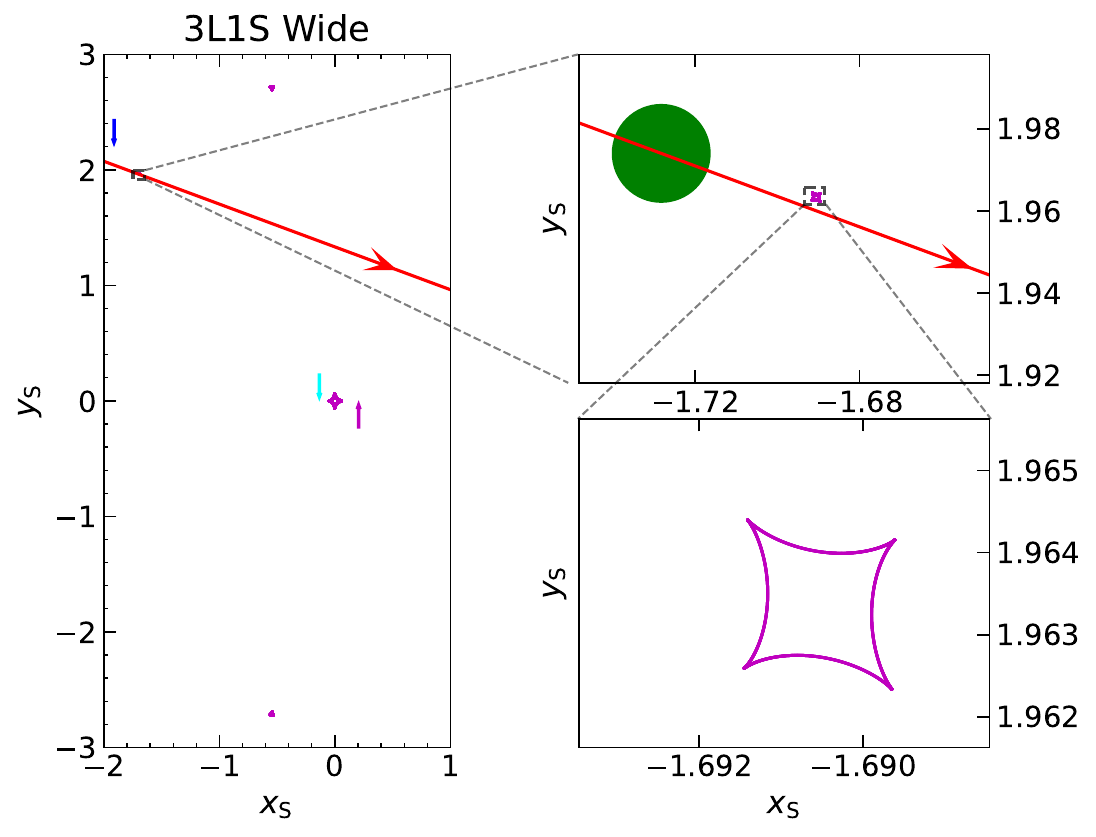}
    %\caption{Caustic geometries of the four static 2L1S solutions. The locations of the host star and planet are indicated by the blue arrows. The magenta lines show the caustic structures, the, the red lines represent the source trajectory, and the red arrows indicate the direction of the source motion. The radii of the green dots represent the best-fit normalized source radius, $\rho$, of each solution.}
    \caption{Caustic geometries of the 3L1S models. The locations of the host star, the planet and the third body are indicated by cyan, blue, and magenta arrows, respectively.}
    \label{fig:cau3}
\end{figure}

We adopt the binary superposition method \citep{Han2001, Han2005} to search for the 3L1S models (see the Appendix of \citealt{OB191470} for the detailed procedures). We adopt the contour integration code \citep{Kuang2021} to calculate the 3L1S magnification. We exclude the data during $7809.4 < \hjd < 7811$ (i.e., the signature of the planet, $M_2$) and conduct a 2L1S grid search, which consists of 41 values of $\log s_3$ equally spaced between $-1.00$ and $1.00$, 21 values of $\log q_3$ equally spaced between $-2.0$ and $0.0$, and 24 values equally spaced between $0^{\circ}\leq \psi < 360^{\circ}$. We fix $\rho = 0$ because the long-term residuals show no caustic-crossing features. We use MCMC to search for the minimum $\chi^2$ and allow ($t_0$, $u_0$, $\te$, $\psi$) to vary. We obtain three local minima on the ($\log{s_3}$, $\log{q_3}$) plane, with ($\log{s_3}$, $\log{q_3}) = (-0.5, 0.0)$, ($-0.15, -0.7$), and ($0.65, -0.8$), respectively. We then refine the three local minima with the MCMC method by setting all parameters as free, and the first local minimum is favored by $\Delta\chi^2 = 10$ and $\Delta\chi^2 = 5$ compared with the second and third local minima, respectively. The purpose of the 3L1S modeling is to investigate whether a 3L1S model can fit the observed data instead of finding all of the models, so we only adopt the parameters of the first local minimum for the binary superposition method. Combining it with the 2L1S static parameters in Table \ref{tab:2L1S}, we also find three models, which are labeled as ``3L1S Close Outer'', ``3L1S Close Inner'' and ``3L1S Wide''\footnote{If one assumes circular orbits with the ratio of radii equal to the ratio of observed projected separations, then the two ``Close'' solutions are dynamically unstable according to conditions presented by \cite{1999HolmanStable}.}. Figure \ref{fig:cau3} shows the caustics and source trajectories of these models, and their parameters are presented in Table \ref{tab:3L1S}. We also conduct a Bayesian analysis following the procedure of Section \ref{lens}, and the only exception is for that we first estimate the primary lens with $\te$ and $\thetae$ by scaling by a factor of $\sqrt{1+q_2+q_3}$ smaller than the values from the 3L1S models. Table \ref{tab:phy} shows the Bayesian results. 

Compared to the best 2L1S xallarap model (``Wide A''), the ``3L1S Close Outer'' model is excluded by $\Delta\chi^2=50.7$ from the light-curve analysis. For the ``3L1S Close Inner'' model, it is disfavored from both the light-curve analysis (11.3) and the Galactic-model likelihood (10.8). The Bayesian analysis suggest a nearby system with two M dwarfs, with an estimated brightness of $I \sim 17.5$ magnitude, which is significantly brighter than the allowed blended flux, so we can also exclude the ``3L1S Close Inner'' model. However, the best-fit 3L1S model, the ``3L1S Wide'' model, is only disfavored relative to the 2L1S xallarap model ``Wide A'' by $\Delta\chi^2 = 6.2$, so we cannot distinguish between the 2L1S xallarap model and the 3L1S model. 

%%The two solutions have similar caustic geometries and source trajectories. Therefore, we take the first solution as the initial parameters of the extra lens component, with ($t_0, u_0, \te, \log s, \log q, \alpha, \log\rho)$ $=$ $(7884.05, 1.49, 29.39, -0.51, -0.16, 341.86^\circ, -3.57$). 

\subsection{Binary-Lens Binary-Source Model}\label{sec:2L2S}

Second, we add a source companion to the static 2L1S model to fit out the long-term residuals, and now the model is a binary-lens binary-source model (2L2S). Different from the 2L1S xallarap model, here we consider the flux and magnification from the source companion but ignore the orbital motions of the two sources. 

Similar to the 1L2S model, the total magnification of a 2L2S model is the superposition of the 2L1S magnification of the two sources. The 2L2S model has an identical definition of the total magnification $A_\lambda$ and the flux ratio $q_{f,\lambda}$ as in Equation (\ref{equ:total_mag}) and (\ref{equ:bright_ratio}). To obtain an initial guess for the secondary source, we first fit a 1L2S model with the planetary anomaly ($7809.4 < \hjd < 7811$) removed. There are no useful constraints on the scaled radius for the secondary source, $\rho_2$, so we adopt a point-like secondary source. Then, we add the binary-lens parameters ($\rho$, $t_{0, \rm pl}$, $u_{0, \rm pl}$, and $t_{\rm E, pl}$) of the 2L1S static models and search for the best-fit models using the MCMC. We also conduct a Bayesian analysis for the 2L2S models.

Table \ref{tab:2L2S} presents the parameters from the MCMC, and Table \ref{tab:phy} shows the Bayesian results. The Inner and Outer models for the Wide topology also merge into one. Compared to the best 2L1S xallarap model (``Wide A''), the ``2L2S Close Outer'' model can be ruled out by $\Delta\chi^2=45.0$. Both the ``2L2S Close Inner'' and ``2L2S Wide'' models are disfavored by $\Delta\chi^2 \sim 20$, indicating that they are significantly disfavored but cannot be fully excluded. 

The two sources have the similar colors, i.e., $q_{{f,V}}/q_{{f,I}} \sim 1$. Different from the very low flux ratio in the $I$ band for the 1L2S models, $q_{f,I} \sim 10^{-3}$, the secondary source is only $\lesssim 2.5$ magnitude fainter than the primary source. According to the CMD (Figure \ref{fig:cmd}), the putative secondary source is a typical bulge main-sequence star or sub-giant, so we cannot rule out the 2L2S models by the color argument.

\begin{table}[htb]
    \renewcommand\arraystretch{1.25}
    \centering
    \caption{Lensing Parameters for the 2L2S Models}
    \begin{tabular}{c|c c c c}
    \hline
    \hline
    Parameters & Close Inner & Close Outer & Wide \\
    \hline
    $\chi^2$/dof & $\mathbf{10089.9/10682}$ & $10116.7/10682$ & $10093.0/10682$\\
    \hline
    $t_{0,1}$ & $\mathbf{7884.82_{-0.74}^{+0.72}}$ & $7888.18_{-0.59}^{+0.49}$ & $7892.27_{-1.96}^{+1.70}$\\
    $u_{0,1}$ & $\mathbf{1.244_{-0.070}^{+0.048}}$ & $1.317_{-0.033}^{+0.020}$ & $1.292_{-0.079}^{+0.046}$\\
    $\te$ (days) & $\mathbf{32.19_{-0.75}^{+1.06}}$ & $30.38_{-0.49}^{+0.52}$ & $30.17_{-0.68}^{+1.42}$\\
    $t_{0,2}$ & $\mathbf{7866.0_{-1.1}^{+1.0}}$ & $7867.7_{-0.6}^{+0.7}$ & $7870.2_{-1.1}^{+0.9}$\\
    $u_{0,2}$ & $\mathbf{0.687_{-0.077}^{+0.051}}$ & $0.851_{-0.033}^{+0.030}$ & $0.974_{-0.082}^{+0.057}$\\
    $\rho_1$ ($10^{-2}$) & $\mathbf{0.100_{-0.033}^{+0.183}}$ & $0.038_{-0.024}^{+0.306}$ & $0.954_{-0.722}^{+0.159}$\\
    $q_{{f,I}}$ & $\mathbf{0.105_{-0.031}^{+0.029}}$ & $0.243_{-0.035}^{+0.033}$ & $0.629_{-0.197}^{+0.212}$\\
    $q_{{f,V}}$ & $\mathbf{0.115_{-0.041}^{+0.032}}$ & $0.250_{-0.044}^{+0.058}$ & $0.699_{-0.216}^{+0.271}$\\
    $t_{\rm 0,pl}$ & $\mathbf{7814.01_{-0.22}^{+0.25}}$ & $7806.56_{-0.28}^{+0.34}$ & $7810.46_{-0.02}^{+0.03}$\\
    $u_{\rm 0,pl}$ & $\mathbf{0.177_{-0.013}^{+0.014}}$ & $-0.308_{-0.031}^{+0.030}$ & $0.005_{-0.002}^{+0.002}$\\
    $t_{\rm E,pl}$ (days) & $\mathbf{-1.19_{-0.09}^{+0.08}}$ & $-1.70_{-0.12}^{+0.14}$ & $0.25_{-0.03}^{+0.02}$\\
    $I_{\rm S, total}$ & $\mathbf{17.374_{-0.063}^{+0.082}}$ & $17.306_{-0.034}^{+0.081}$ & $17.325_{-0.047}^{+0.080}$\\
    \hline
    $q_{{f,V}}/q_{{f,I}}$ & $\mathbf{1.07_{-0.19}^{+0.18}}$ & $1.04_{-0.14}^{+0.16}$ & $1.13_{-0.12}^{+0.15}$\\
    $s$ & $\mathbf{0.3382_{-0.0071}^{+0.0103}}$ & $0.3144_{-0.0046}^{+0.0046}$ & $3.2972_{-0.1325}^{+0.1029}$\\
    $q$ ($10^{-4}$) & $\mathbf{13.60_{-1.57}^{+1.90}}$ & $31.46_{-5.59}^{+5.15}$ & $0.71_{-0.16}^{+0.12}$\\
    $\alpha$ (deg) & $\mathbf{32.9_{-0.6}^{+0.6}}$ & $20.5_{-0.7}^{+0.8}$ & $205.5_{-0.9}^{+0.9}$\\
    \hline
    \hline
    \end{tabular}
\label{tab:2L2S}
\end{table}

\subsection{Implications: Dimension-Degeneracy Disasters}\label{sec:imp}

Together with the 3L1S and 2L2S models, we have respectively investigated four effects to fit out the long-term residuals. They are parallax, xallarap, an additional lens, and an additional source. From the perspective of light-curve analysis alone, the four effects all can fit the light curve well and we cannot exclude any effect. To the best of our knowledge, before this case, only \cite{Yang_TLC} respectively explored all four of these effects for a single event. In that case, two of the effects (parallax or adding a source) were excluded because they  cannot fit the light curve ($\Delta\chi^2 \gtrsim 50$). Therefore, for the first time a severe degeneracy between the four effects was found in a real case. In addition, \cite{Yang_TLC} investigated an anomaly from a 1L1S model, while we analyzed an anomaly to a 2L1S model, which significantly added difficulties in analysis and computation. 

%Although we finally rule out the parallax effect due to the very low probability of the large parallax value from a Bayesian analysis based on a Galactic model, for the cases with weaker high-order signals, the parallax values can be smaller and it might be hard to exclude the parallax effect. 

In principle, we can further fit the light curve by combining any two of the four effects, e.g., the 3L1S parallax model and the 2L2S xallarap model. For example, \cite{OB180532} tried the 3L1S parallax model, the 2L2S parallax model and a three-effect model (the 2L2S parallax xallarap model) for the planetary event OGLE-2018-BLG-0532. We stopped at the one-effect models because the analysis that has been done in this paper was already one of the most complicated analysis over all published microlensing events and there are no clear prospects for further investigations. That is, it is unlikely that further investigations can break the Close/Wide degeneracy or change the nature of the planet, i.e., a Jovian-mass planet at $\sim 1$ au or a super-Earth-mass to Neptune-mass planet at $\sim 6$ au (Table \ref{tab:phy}). All of the four effects can fit the long-term residuals well and thus more dimensions would probably only loosen the constraints on each effect and decrease the $\Delta\chi^2$ between the models. Of course, one might argue that the parallax effect exists in any model anyway because of Earth's orbital motion, but a physical parallax effect for the present case, i.e., for a short ($\te \sim 30$ days) and low-magnification ($u_0 \sim 1.3$) event, would not produce a detectable signature and thus not significantly affect the parameters of the other three effects. This is different from the \cite{OB180532} case, for which $\te \sim 140$ days and thus the parallax effect played an important role. 

Nevertheless, the degeneracy leads to concerns for several studies. First, the detection of isolated stellar-mass black holes from the annual microlensing parallax may need to consider the influences from the other three effects. For example, the existence of the third lens in the event, KMT-2020-BLG-0414, significantly affected the measurements of the annual microlensing parallax, and being unaware of the third lens would have led to a misjudgment on the primary lens \citep{KB200414}. Because the satellite microlensing parallax \citep{1966MNRAS.134..315R,1994ApJ...421L..75G} is measured by the differences between the light curves observed from Earth and a satellite or two satellites \citep{Zhu2017K2Spitzer}, which are overwhelming compared to the differences between the annual parallax and the other effects, maybe the only robust way to make industrial-scale detections of isolated stellar-mass black holes is the satellite microlensing parallax \citep{Gould_BH}. Second, a systematic study of the stellar binaries may need to analyze only the events with clear caustic-crossing features because they must be caused by adding a lens, instead of parallax, xallarap or adding a source. Our long-term residuals are not rare for stellar binary-lens events, which can be inferred from the three local minima that we obtained from the 2L1S grid search in Section \ref{sec:3L1S}. Including and analyzing such non-caustic-crossing features with the other three effects would be too time-consuming (and thus painful) to form a large statistical sample. Note that such pains would not be reduced for a systematic study of planets in binary systems. Although the rate of ambiguous and unambiguous planetary events is only $\sim 2\%$ for the current KMTNet survey, adding the four effects to the 2L1S models is at least one order of magnitude more complicated than adding them to the 1L1S models, and the planetary rate will be higher for the future space-based microlensing projects (Roman, \citealt{MatthewWFIRSTI}; Earth 2.0, \citealt{ET}; CSST, \citealt{CSST_Wei}) because of their stable photometry and complete coverage.

Another challenge is how to identify the necessity of trying these effects. The current ``standard'' light-curve analysis only tries the parallax effect first and only investigates more effects if the resulting parallax is detected or suspicious (e.g., by further checking for xallarap). Our case followed this procedure exactly. However, is it possible that this ``standard'' procedure cannot identify the clues of the other three effects even when they are detectable? It is alert that without the four effects the ``Close Inner'' model is favored by $\Delta\chi^2 \geq 40$ over the other static models, which is significant enough to rule out the other static models, but in fact the Wide topology is preferred with considering these effects. Therefore, being unaware of these effects might lead to a wrong conclusion about the nature of the lens system (e.g., the planet). 

Finally, our analysis stopped at the one-effect models, but there is currently no clear endpoint of the analysis. One can argue that each effect is possible and thus the microlensing modelers should explore as further as they can. Note that in the above discussion, we have not included the lens orbital motion effect. For our event, due to the low mass ratio of the secondary lens and the short-lived planetary signal, the orbital motion of the lens system has almost no effect on the light curve, but in some cases, the lens orbital motion effect can be degenerate with other effects (e.g., \citealt{MB09387,OB09020}). In the most extreme case, there are more than 20 parameters if we consider all of the effects, which is intolerable in terms of computational resources and is likely unnecessary because the observed data cannot simultaneously provide useful constraints on all of the effects.

\section{Summary}\label{sum}

We have conducted an analysis of the microlensing planetary event, \event. The planetary signal was observed by both the KMTNet and OGLE surveys and identified by the KMTNet AnomalyFinder algorithm. To fit the planetary signal, we first tried the static 2L1S models and found two families of models: one with a low-mass-ratio wide-orbit ($s > 2$) planet (exhibiting the inner-outer degeneracy) and another (with two degenerate solutions) with a larger-mass ratio planet in a close ($s<1$) orbit. However, there were still long-term residuals in the light curve. Thus, we further investigated the light curve by adding the microlensing parallax (the 2L1S parallax models), the microlensing xallarap (the 2L1S xallarap models), an additional lens (the 3L1S models), and an additional source (the 2L2S models). Adding these effects resulted in additional degenerate solutions and further uncertainty in the mass-ratio of the planet. We also considered 1L2S with extreme xallarap models, but these turned out to be unphysical. With a Bayesian analysis, the 2L1S parallax models are excluded but the 3L1S and 2L2S models still survive. The 2L1S xallarap wide-orbit ``A'' model provides the best fit, with $q = 0.98 \times 10^{-4}$ and $s = 2.45^{+0.40}_{-0.30}$, and its mass ratio is lower than the previously lowest $q$ ($2.1 \times 10^{-4}$) for planets with $s > 2$ \citep{OB180383}. However, we cannot rule out the close-orbit models with $q \sim 10^{-3}$ and $s \sim 0.35$. The wide-orbit models all contain a super-Earth-mass to Neptune-mass planet at a projected planet-host separation of $\sim 6$ au and the close-orbit models all consist of a Jovian-mass planet at $\sim 1$ au (Table \ref{tab:phy}).

All of the four effects can fit the light curve well and result in variations in the mass-ratio between solutions that exceed the uncertainties for a given solution. Furthermore, the preference for the larger-mass ratio, close solution versus the lower-mass ratio, wide solution changes depending on which effects are included. This creates a ``curse of dimensionality'' in analyzing microlensing light-curves and interpreting their planets. We call for more studies investigating these issues from theoretical, simulation, and statistical perspectives. These studies are urgently needed because the Roman and Earth 2.0 teams are currently building their modeling pipelines.

\bigskip

\acknowledgments
R.Z., W.Zang, S.M., H.Y., R.K., J.Z., and W.Zhu acknowledge support by the National Natural Science Foundation of China (Grant No. 12133005). Work by R.P. and J.S. was supported by Polish National Agency for Academic Exchange grant ``Polish Returns 2019.'' W.Zang acknowledges the support from the Harvard-Smithsonian Center for Astrophysics through the CfA Fellowship. This research has made use of the KMTNet system operated by the Korea Astronomy and Space Science Institute (KASI) and the data were obtained at three host sites of CTIO in Chile, SAAO in South Africa, and SSO in Australia. Data transfer from the host site to KASI was supported by the Korea Research Environment Open NETwork (KREONET). This research was supported by the Korea Astronomy and Space Science Institute under the R\&D program (Project No. 2023-1-832-03) supervised by the Ministry of Science and ICT. This research uses data obtained through the Telescope Access Program (TAP), which has been funded by the TAP member institutes. Work by J.C.Y. and I.-G.S. acknowledge support from N.S.F Grant No. AST-2108414. Work by C.H. was supported by the grants of National Research Foundation of Korea (2019R1A2C2085965 and 2020R1A4A2002885). Y.S. acknowledges support from BSF Grant No. 2020740. The authors acknowledge the Tsinghua Astrophysics High-Performance Computing platform at Tsinghua University for providing computational and data storage resources that have contributed to the research results reported within this paper. 

\software{
EMCEE \citep{emcee}, 
MulensModel \citep{2019A&C....26...35P}, 
cmplx\_roots\_sg \citep{2012arXiv1203.1034S},
VBBL \citep{Bozza2010,Bozza2018}, triplelens \citep{Kuang2021}
}

\bibliography{Zang.bib}

\end{document}